%% file: final.tex
\documentclass[superscriptaddress,prd,twocolumn,aps,amssymb,nofootinbib]{revtex4}

\usepackage{latexsym}
\usepackage{graphicx}
\usepackage{amssymb}
\usepackage{lscape}
\usepackage{dcolumn}
\usepackage{bm}

\newcommand{\be}{\begin{equation}}
\newcommand{\ee}{\end{equation}}
\newcommand{\bea}{\begin{eqnarray}}
\newcommand{\eea}{\end{eqnarray}}

\newcommand{\OOD}[1]{\partial_{#1}}

\newcommand{\TD}[2]{\tilde{#1}_{#2}}
\newcommand{\TU}[2]{\tilde{#1}^{#2}}
\newcommand{\TUU}[3]{\tilde{#1}^{#2 #3}}
\newcommand{\TDD}[3]{\tilde{#1}_{#2 #3}}
\newcommand{\TOD}[2]{( \tilde{\mathbb L} \tilde{\omega} )_{#1 #2}}
\newcommand{\TOU}[2]{( \tilde{\mathbb L} \tilde{\omega} )^{#1 #2}}

\newcommand{\LL}{\mathbb L}

\newcommand{\OO}{\partial}

\newcommand{\GA}{\alpha}
\newcommand{\GB}{\beta}
\newcommand{\GG}{\gamma}
\newcommand{\GD}{\delta}

\newcommand{\GL}{\lambda}
\newcommand{\GR}{\rho}

\newcommand{\GC}{\psi}

\newcommand{\GO}{\omega}
\newcommand{\GZ}{\zeta}
\newcommand{\GP}{\phi}

\newcommand{\GJ}{\theta}

\input kigou.tex

\begin{document}

\title{Numerical method for binary black hole/neutron star initial data: Code test}

\author{Antonios A. Tsokaros}
\affiliation{Department of I.C.S.E., University of Aegean, Karlovassi 83200, Samos, Greece} 

\author{K\=oji Ury\=u}
\affiliation{Department of Physics, University of Wisconsin-Milwaukee, P.O. Box 413,  
Milwaukee, WI 53201}

\date{\today}

\begin{abstract}
A new numerical method to construct binary black hole/neutron 
star initial data is presented.  The method uses three 
spherical coordinate patches; Two of these are centered at 
the binary compact objects and cover a neighborhood of each 
object; the third patch extends to the asymptotic region.
As in the Komatsu-Eriguchi-Hachisu method, nonlinear elliptic field equations are 
decomposed into a flat space Laplacian and a remaining nonlinear 
expression that serves in each iteration as an effective source. 
The equations are solved iteratively, integrating a Green's function 
against the effective source at each iteration. 
Detailed convergence tests for the essential part of the code 
are performed for a few types of selected Green's functions 
to treat different boundary conditions.  Numerical computation 
of the gravitational potential of a fluid source, and a toy model 
for a binary black hole field are carefully calibrated with the 
analytic solutions to examine accuracy and convergence of the new code.  
As an example of the application of the code, an initial data set for 
binary black holes in the Isenberg-Wilson-Mathews formulation is presented, 
in which the apparent horizons are located 
using a method described in Appendix A.  
\end{abstract}

\maketitle

\section{Introduction}

Inspiral and merger simulations to produce accurate 
gravitational waveforms are essential for constructing 
waveform templates for analysis of data from 
laser interferometric detectors.  
The ground based interferometers, such as Advanced LIGO or LCGT 
may detect gravitational waves from the inspiral of $M\sim 10 M_\odot$ 
binary black holes within $z\sim 4$, 
while space based interferometric detectors such as 
LISA or DECIGO may detect the inspiral of $10^6 M_\odot$ 
supermassive binary black holes, and may discover 
intermediate mass binary black holes with $M\sim 10^3M_\odot$.  

Initial data set for binary black holes 
with a variety of black hole parameters, such as binary 
mass ratio, and black hole spins, or the binary black hole-neutron 
star data will become more useful considering the remarkable 
progress made recently for the inspiraling binary black holes 
simulations up to a few orbits near the innermost stable circular orbits  
\cite{BTJ2004,ABDGo2005,Pretorius2005,NASA2006,BROWN2006,Diener2006,HSL2006,
Brug2006}.  
Several groups have been achieved to construct binary black hole 
initial data successfully \cite{Baumg2000,PTC2000,MM2000,GGB2002,
Cook2002,PCT2002,TBCD2003,Ansorg,YCST2004,Hannam2005}, 
(for earlier works, see \cite{Cook2000}).  

In this paper, we introduce a new numerical method suitable 
for computing accurate initial data sets for binary black 
holes, black hole-neutron star binaries, and binary neutron 
star systems.  
Initial data sets of these kinds are calculated from 
the Einstein equation written in the form of nonlinear 
elliptic equations for metric components. Each equation 
can be written as a Poisson equation with a nonlinear source.  
Our new Poisson solver is patterned after the Komatsu-Eriguchi-Hachisu (KEH) 
method \cite{KEH89},
widely used to compute rotating neutron stars and more 
recently to compute, binary neutron stars \cite{USE00}. 
The KEH method uses Green's formula to write the field 
equations in equivalent integral forms and iteratively 
solve them using spherical coordinates and angular harmonics. 
The set of equations is discretized by a standard finite 
difference scheme.  

To extend the method to handle our wider class of binary 
configurations, we introduce three spherical coordinate patches. 
Two are centered at each hole and extend outward 
to a finite radius, larger than the gravitational radius but small 
enough that the two coordinate patches do not overlap.  A third coordinate 
patch, covering the rest of a spacelike hypersurface, extends 
to the the asymptotic region and overlaps each of the 
other two patches.   
There are two important features of our new code: 
(1) The number of multipoles in the coordinate patch centered 
at the orbital center can be reduced to $\ell \alt 10$ 
since the sizes of patches for compact objects is extended to 
about a half of the binary separation.  
(2) The data between those patches are communicated 
only at the boundary of those patches to minimize the amount of 
data to interpolate from one to the other.  
These novel features result in an efficient code that retains 
high resolution even near the compact objects where the field is strong.  
The method has two additional significant advantages: Coding is 
relatively simple, and the iteration converges robustly.  

This paper is organized as follows.
In section \ref{sec:method}, after briefly reviewing the 
initial value formulation, we introduce our choice of coordinate 
systems and the formulation of our Poisson solver.
Formulas for multipole expansion of Green's function used in 
the Poisson solver are described in Appendix \ref{sec:ApGreen}.  
In section \ref{sec:coding}, we describe our numerical methods, 
including finite differencing and our iteration procedure. 
In the relating Appendices \ref{sec:Aptoy} and \ref{sec:APconv}, 
the convergence of iteration is discussed further.  
The results of detailed convergence tests 
are presented in \ref{sec:codetest}, and an example of 
binary black hole initial data is displayed in 
\ref{sec:inidata}.  
The concrete form of the nonlinear elliptic 
equations for the initial value problem is given in 
Appendix \ref{sec:eqs}, and a method 
to locate the apparent horizons in the initial data is 
described in Appendix \ref{sec:aphzn}.

\section{Method for binary black hole/neutron star initial data}
\label{sec:method}

Our new numerical method is applicable for various formulations 
including spatially confomal flat initial data \cite{SourcesOfGR}, 
the Isenberg-Wilson-Mathews (IWM) formulation \cite{ISEN78,WMM959,BA97}, 
and waveless approximation \cite{SUF04}.  
We introduce the IWM formulation used for a test calculation 
of our new code. In this formulation, four constraints and the spatial 
trace of the Einstein equation are solved, after choosing  
the trace of the extrinsic curvature and the conformal three metric.  
We then explain the choice of coordinates and the form of 
Green's functions used in our version of the KEH method.

\subsection{Formulation}

We consider a globally hyperbolic spacetime $\cal M$,  
foliated by a family of spacelike hypersurfaces $\Sigma_t$.  
The binary black hole initial data is constructed on 
a slice $\Sigma=\Sigma_0$.  The unit future-pointing normal to 
$\Sigma_t$ will be denoted by $n_\alpha=-\alpha\na_\alpha t$ and 
the metric is written in 3+1 form,
\bea
ds^{2} & = & g_{\mu\nu}dx^{\mu}dx^{\nu} \nonumber \\
       & = & -\GA^{2}dt^{2}+\GG_{ij} (dx^{i}+\GB^{i}dt) (dx^{j}+\GB^{j}dt), 
\eea
in a chart $\{t,x^i\}$, 
where $\GG_{ab}(t)$ is the 3-metric on $\Sigma_t$, $\alpha$ the lapse 
and $\beta^a$ the shift.  The 3-metric $\GG_{ab}$ is induced by 
the projection tensor to the hypersurfaces $\Sigma_t$
\be
\GG_{\GA\GB}=g_{\GA\GB}+n_{\GA}n_{\GB}.  
\ee 

The extrinsic curvature of the foliations is defined by 
\be
K_{\GA\GB}=- \frac{1}{2} \Lie_{n} \GG_{\GA\GB}, 
\ee 
where $K_{\GA\GB}$ satisfies $K_{\GA\GB} n^\alpha=0$.  
With the spatial indices, the spatial tensor $K_{\albe}$ 
is written 
\be
K_{ab}
= -\frac1{2\alpha} \left(\pa_t\gmabd
-\Lie_\beta\gmabd \right).
\label{eq:Kab}
\ee
Denoting the tracefree part of $K_{ab}$ by $A_{ab}$ 
and its trace by $K:=K_a{}^a$, we have
\be
A_{ab} = K_{ab} - \frac13 \gmabd K,
\label{eq:decomposeK}
\ee
and 
\be
A_{ab}= - \frac{1}{2} 
\left(\Lie_{n} \gmabd - \frac13\gmabd\gamma^{cd}\Lie_{n} \gamma_{cd}\right).  
\label{eq:Aab}
\ee
$\Lie_n$ operating to the spatial metric is understood as 
$\Lie_n = \frac1{\alpha}(\pa_t - \Lie_\beta)$, where the 
$\Lie_\beta$ is the Lie derivative defined on $\Sigma$.  

Projecting one index of the  Einstein equation 
normal to the hypersurface $\Sigma$, $G_{\albe}n^\alpha=0$, 
yields the Hamiltonian and momentum constraint equations 
\beqn
&&2(\Gabd-8\pi T_{\albe})n^\alpha n^\beta 
\nonumber\\&&\qquad
= R - K_{ab} K^{ab}+K^{2} - 16\pi\rho_{\rm H}=0,  \\   
&&(\Gabd-8\pi T_{\albe})\gamma^{\beta a} n^\alpha 
\nonumber\\&&\qquad
= -D_{b}(K^{ab}-\GG^{ab}K) + 8\pi j^{a}= 0,   
\eeqn
where $D_a$ is the covariant derivative on $\Sigma$ 
associated with the three metric $\gamma_{ab}$.  
To satisfy the Hamiltonian constraint on a slice $\Sigma$, 
we introduce a conformal decomposition of the spatial metric, 
$\gmabd=\psi^4\tgmabd$, and solve the Hamiltonian constraint
for the conformal factor $\psi$.  This prescription leaves the 
conformal three metric $\tgmabd$ unspecified.  

Separating out the trace $K$, and substituting 
Eqs.~(\ref{eq:decomposeK}) and (\ref{eq:Aab}) 
in the momentum constraint results in an elliptic equation 
for the shift $\beta^a$. The trace, $K$, remains unspecified.  
The spatial trace of the Einstein equation, 
\beqn
&&(\Gabd-8\pi T_{\albe})\gamma^{\albe} 
\nonumber\\
&&= 2\Lie_n K -\frac12(R+K^2+3K_{ab}K^{ab})
+\frac2\alpha D_a D^a\alpha - 8\pi S
\nonumber\\
&&=0, 
\eeqn
can be written as an elliptic equation for the lapse $\alpha$,
once one restricts $\partial_t K$ (e.g., by setting 
to zero $\partial_t K$ or the derivative of $K$ along a 
helical Killing vector). 

In Appendix \ref{sec:eqs}, we show the explicit form of the elliptic equations
with nonlinear source for the constraints and the spatial
trace of the Einstein equation.

\subsection{Inversion of the Laplacian: Poisson solver}
\label{sec:Poisol}

In each component of the field equations a second-order elliptic operator acts on 
one metric potential.  By separating out a flat Laplacian, 
we write each field-equation component in the form 
\beq
\na^2 \Phi = S,
\label{eq:Poisson}
\eeq
where $\Phi$ represents a metric potential  
on a slice $\Sigma$. The effective source $S$ involves second 
derivatives of the metric, but a convergent iteration is possible because 
they occur in expressions that are $o(r^{-3})$ near spatial infinity.      
The flat Laplacian $\na^2$ is separated in spherical coordinates and inverted by 
a Poisson solver, and the nonlinear equation (\ref{eq:Poisson}) 
is solved iteratively.    
Our choice of the Poisson solver is to use the Green's formula, 
an integral form of Eq.~(\ref{eq:Poisson}).  
Using the Green's function of the Laplacian 
\beq
\na^2 G(x,x') = -4\pi\,\dl(x-x'), 
\label{eq:Greeneq}
\eeq
where $x$ and $x'$ are positions, $x,x'\in V \subseteq \Sigma$, 
the Green's formula is written 
\beqn
\Phi(x)&=& -\frac1{4\pi}\int_{V} G(x,x')S(x') d^{3}x' 
\nonumber \\
&&
+ \frac{1}{4\pi} \int_{\OO V} \left[G(x,x')\na'^a \Phi(x')\right.
\nonumber \\
&&\qquad\qquad
\left. - \Phi(x')\na'^a G(x,x') \right]dS'_a.  
\label{eq:GreenIde}
\eeqn
This formula is valid for any connected space $V$ as long as each 
term is integrable.
The function $G(x,x')$ is a sum of a Green's function without boundary 
and a homogeneous solution $F(x,x')$ to the Laplace equation, 
\beq
G(x,x') = \frac1{\left| {x}-{x}' \right|} + F(x,x'), 
\label{eq:Green}
\eeq
which satisfy
\beqn
\na^2 \frac1{\left| {x}-{x}' \right|} &=& -4\pi\,\dl(x-x'), \\
\label{eq:GreenNBeq}
\na^2 F(x,x') &=& 0.  
\label{eq:Greeneq2}
\eeqn
Eq.~(\ref{eq:GreenIde}) is a formal solution to the Poisson equation 
(\ref{eq:Poisson}) for any $G(x,x')$ that satisfies 
Eq.~(\ref{eq:Greeneq}), 
even if the source $S$ depends on the field $\Phi$ nonlinearly. 
Thereby, the elliptic equation with a nonlinear source 
can be solved iteratively using Eq.~(\ref{eq:GreenIde}).  
We call this iteration the KEH iteration hereafter.  

With the surface term included, Eq.~(\ref{eq:GreenIde}) is an identity, 
valid for any choice of Green's function.  Requiring convergence of 
the KEH iteration, however, imposes the following key restrictions on that choice:   
For solving a Dirichlet problem, no multipole 
component of $\nabla'^a G(x,x')$ can vanish   
on the entire boundary; similarly, for solving a Neumann problem, 
no multipole component of $G(x,x')$ can vanish on the entire boundary.  
As long as the Green's function satisfies this restriction, 
it is not necessary to construct $F(x,x')$ appropriate for each 
boundary condition; for example, the Green's function without boundary 
term $1/\left| {x}-{x}' \right|$ can be used for the Neumann problem.  

A reason to use the Poisson solver Eq.~(\ref{eq:GreenIde}) 
is tied to the facts that the spherical coordinates $(r,\theta,\phi)$
are suitable for constructing numerical domains for binary black holes
and neutron stars, 
and that one can use a multipole expansion of the Green's function 
in these coordinates.  The Poisson solver turns out to be simple for 
coding, CPU inexpensive and accurate as shown in Sec.\ref{sec:codetest}.  
In the rest of this section, we introduce numerical domains for 
the binary black holes, and discuss the choice of the Green's 
function.  

\begin{figure}
  \includegraphics[width=3.1in,clip]{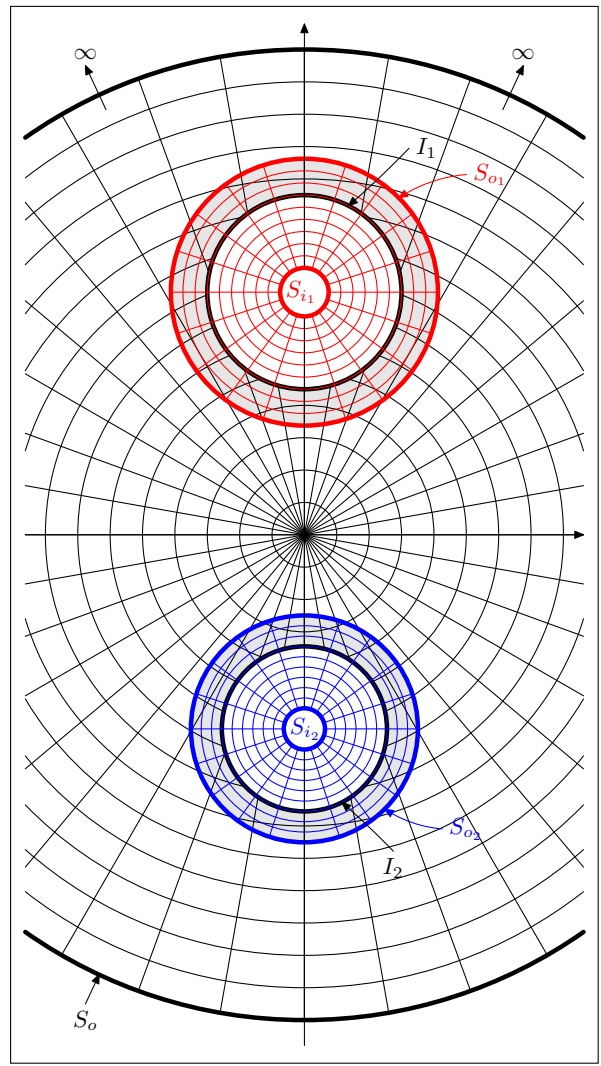}
  \caption{The computational domain. 
  One central grid and two black hole grids with excised regions.   }
  \label{FigTotalCG}
\end{figure}

\subsection{Construction of the computational domain}

To describe the binary black hole/neutron star data, we introduce 
three spherical domains.  Two small domains are centered 
at the two compact objects surrounding them, while the third domain 
partly covers the first two and extends to the asymptotic region.  
The large domain is not necessarily positioned at the center of mass 
of the two compact objects.  
These domains are shown schematically in Fig.~\ref{FigTotalCG}.  
In this section, we consider the binary black hole case as an example, 
and refer to the domains around the compact objects as the 
black hole coordinate system (BHCS)
and the third one as the central coordinate system (CCS).  

CCS extends to the asymptotic region; practically 
the radius $r$ of the sphere $S_{o}$ is set 
large enough that the multipoles of order $r^{-2}$
and higher are negligible.  
It excludes the interiors of the two spheres $I_{1}$ and $I_{2}$, which 
are centered at each black hole, and whose  
radii are taken larger than the gravitational radius of each hole 
but not as large as to intersect each other.  
Therefore, in the domain of CCS, Eq.~(\ref{eq:GreenIde}) 
involves (1) the surface integrals over a large sphere 
$S_{o}$ where the asymptotic condition of 
each field variable is imposed, (2) the interfaces 
$I_{1}$ and $I_{2}$, and (3) the volume integral of 
the source term $S$ between these three spheres.  

The first black hole computational domain BHCS-1 extends from 
a sphere $S_{i_{1}}$ to $S_{o_{1}}$, both centered at the black hole,  
and the second one BHCS-2 from $S_{i_{2}}$ to $S_{o_{2}}$.  
The region inside $S_{i_{1}}$ and $S_{i_{2}}$ is excised from
the computational domain. The code allows the option of dispensing 
with the inner boundary, for a black hole; and for a neutron star 
the inner boundary is never used (in this case in the BHCS the minimum
value of $r$ is zero).
Note that the spheres $I_1$, $S_{i_{1}}$, and $S_{o_{1}}$ are 
concentric, and the same for $I_2$, $S_{i_{2}}$, and $S_{o_{2}}$.  

The boundaries $I_1$, $I_2$, $S_{o_{1}}$, and $S_{o_{2}}$ are 
introduced to reduce the number of terms in the Legendre expansion of 
the Green's function (\ref{eq:Greenfn}) in CCS.  
Taking the radii of $I_1$ and $I_2$ large enough, 
the contribution of higher multipoles in CCS is included 
in the surface integrals over $I_1$ and $I_2$.  
Because of this, a small number of multipoles, typically 
$\ell\alt 10$, is enough to resolve the volume integral of CCS.  
Thus by increasing the radial resolution in BHCS 
where the metric potentials may vary rapidly, we can compute 
an accurate solution without having a high resolution in CCS.  

Between concentric spheres $I_1$ and $S_{o_{1}}$ of BHCS-1, 
and $I_2$ and $S_{o_{2}}$ of BHCS-2, we reserve overlapping 
regions, which appear shaded in Fig.~\ref{FigTotalCG}.  
These interfaces $I_i$ and $S_{o_{i}}$ ($i=1$ or $2$) 
are not physical boundaries; the boundary conditions 
of the fields are not prescribed there.  Instead, the value of 
the field on $I_i$ is calculated from the field on BHCS, 
and the value of the field on $S_{o_{i}}$ from the field on CCS, 
thus resulting to a smooth potential field 
throughout BHCS and CCS that satisfies the physical 
boundary conditions at the BH boundary and the asymptotic region.  
The significance of the overlap region is to decrease 
the number of iterations to convergence.  A toy model of this 
iteration procedure is explained in Appendix \ref{sec:Aptoy}.

\subsection{Choices for the Green's function}

In Sec.~\ref{sec:Poisol}, we discussed a restriction on the 
choice of Green's function due to the KEH iteration using 
Eq.~(\ref{eq:GreenIde}).  
Any Green's function that meets this restriction 
can be used in the Poisson solver (\ref{eq:GreenIde}).  
The Green's functions are expanded in 
multipoles over the spherical coordinates $(r,\theta,\phi)$
of each domain.  Explicit formulas for the expansions 
are shown in Appendix \ref{sec:ApGreen}.  
In actual numerical computations, the summation of multipoles in $\ell$ 
is truncated at a certain finite number $L$, for which we typically 
choose $L \sim 10$ in order to resolve the deformation of the field $\Phi$.  

For CCS, we choose the Green's function without boundary 
\beq
G^{\rm NB}(x,x')=\frac1{\left|x-x' \right|}, 
\label{eq:GreenNB}
\eeq
which has the simplest form and picks up the contributions 
from the interfaces $I_1$ and $I_2$.  
In the volume integral of Eq.~(\ref{eq:GreenIde}) over the domain 
outside of spheres $I_1$ and $I_2$, and inside of $S_{o}$, 
the function $G^{\rm NB}(x,x')$ is expanded in multipoles over 
the spherical coordinates of CCS.  For the surface integrals on 
$I_1$ and $I_2$, $G^{\rm NB}(x,x')$ is expanded in multipoles 
over the spherical coordinates of BHCS-1 and BHCS-2 respectively.
Therefore, the position $x$ corresponding to each grid point of CCS is
labeled by the spherical coordinates of BHCS, not CCS, in these
surface integrals.

For BHCS, when the excision of the computational domain is not used
at the black hole, $G^{\rm NB}$ will be chosen.  
However when the computational region is excised inside a sphere 
$S_{i_1}$ and $S_{i_2}$,
$G^{\rm NB}$ can not be used for the Dirichlet problem.  
As shown in Appendix \ref{sec:APconv}, the $\ell=0$ component of 
$\na'^a G^{\rm NB}(x,x')$ becomes zero at the inner boundary sphere, 
and hence it can not pick up Dirichlet data there 
during the iteration of Eq.~(\ref{eq:GreenIde}).  
When the black hole boundary condition for a certain field is given 
by Dirichlet data, we choose 
the Green's function for the Dirichlet problem between two concentric 
spheres $G^{\rm DD}$ given in Appendix \ref{sec:APGDD}.  
When Neumann data is imposed at the black hole boundary, 
the Green's function without boundary $G^{\rm NB}$ may be used.  
We also coded the Green's function between two concentric sphere 
$G^{\rm ND}$ for which the Neumann condition is imposed at 
the inner boundary of BHCS $S_{i_1}$ and $S_{i_2}$, and the Dirichlet data at the outer 
boundary of BHCS $S_{o_1}$ and $S_{o_2}$.

\section{method for numerical computings}
\label{sec:coding}

\subsection{Grid spacing}

Hereafter, coordinate labels $(r,\theta,\phi)$ will be used for 
all three spherical coordinate systems, CCS, BHCS-1, and BHCS-2 
unless otherwise stated.  
We introduce three spheres, $S_a$, $S_b$, and $S_c$ at $r=r_a$, $r=r_b$, and 
$r=r_c$ respectively, such that $r_a < r_c < r_b$, in each 
coordinate system.  
The sphere $S_a$ is used as an inner boundary 
for BHCS when the excision boundary is used, which 
corresponds to $S_{i_1}$ and $S_{i_2}$ in Fig.~\ref{FigTotalCG}.  
For CCS or BHCS without excision, the radius $r_a=0$ is understood.  
The sphere $S_b$ is the outer boundary of each coordinate system
that corresponnds to $S_{o}$, $S_{o_1}$, and $S_{o_2}$ in the same figure.  
The sphere $S_c$ is located between $S_a$ and $S_b$ where we change 
the grid spacing in the radial coordinate.  

The code is constructed to handle non-equidistant grid spacing in 
each coordinate grid.  In CCS, the grid starts with equidistant 
spacing from the origin $r=0$ to a sphere $S_c$, and 
from there it becomes non-equidistant with an ever increasing spacing 
up to the outer boundary $S_b$. The black hole grids are equidistant 
from their outer boundaries $S_b$ down to $S_c$, 
and from that point until the inner boundary $S_a$ they become 
non-equidistant with an ever decreasing spacing. 
For the black hole grids, finer grids are adequate to have an accurate 
representation of the rapidly varying fields near the hole, 
while further away, where the potentials are changing slowly, larger 
spacing can be used without compromising accuracy.  

We summarize  common notations for all three coordinate systems as follows;\\
\begin{tabular}{lll}
$r_{a}$ &:& Radial coordinate where each grid starts.                          \\
$r_{b}$ &:& Radial coordinate where each grid finishes.                        \\
$r_{c}$ &:& Radial coordinate between $r_{a}$ and $r_{b}$ where each          \\
&\phantom{:}& grid changes from equidistant to non-equidistant               \\
&\phantom{:}& or vice versa.                                              \\  
$N_{r}$ &:& Total number of intervals $\Dl r_i$ between $r_{a}$ and $r_{b}$. \\
$n_{r}$ &:& Number of intervals $\Dl r_i$ between $r_{a}$ and $r_{c}$.       \\
$n_{v}$ &:& Number of overlapping intervals of BHCS to CCS.                  \\
$N_{\theta}$ &:& Total number of intervals $\Dl \theta_i$ for $\theta\in[0,\pi]$. \\
$N_{\phi}$ &:& Total number of intervals $\Dl \phi_i$ for $\phi\in[0,2\pi]$. \\
$d$ &:& the separation between centers of BHCS and CCS. \\
\end{tabular}  

In particular, we use the following setup for the grid spacings.  \\
\underline{Central Grid}
\begin{eqnarray*}
\Delta r_{i}=\Delta h=\frac{r_{c}-r_{a}}{n_{r}} 
\:\:\: &\textup{for}& \:\:\: 1 \leq i \leq n_{r}
\\
\Delta r_{i+1}=k \Delta r_{i}  \:\:\: &\textup{for}& \:\:\: n_{r} \leq i \leq N_{r}-1 
\end{eqnarray*}
where $k>1$. Then we have
\be
r_{b}-r_{c}=\Delta h \frac{k-k^{N_{r}-n_{r}+1}}{1-k} \:.
\ee
Given $r_{a}$, $r_{b}$, $r_{c}$, $N_{r}$, and $n_{r}$ this is the equation that will give us the spacing factor $k$
for the central grid.  
Note that for the central grid, $r_{a}=0$.
\newline
\underline{Black Hole Grids (I and II)}
\begin{eqnarray*}
\Delta r_{i}=\Delta h=\frac{r_{b}-r_{c}}{N_{r}-n_{r}}  
\:\:\: &\textup{for}& \:\:\: n_{r}+1 \leq i \leq N_{r} 
\\
\Delta r_{i}=k \Delta r_{i+1}  \:\:\: &\textup{for}& \:\:\: 1 \leq i \leq n_{r}   
\end{eqnarray*}
where $k<1$. Then we have
\be
r_{c}-r_{a}=\Delta h \frac{k-k^{n_{r}+1}}{1-k} \:.
\ee

For the angular $\GJ$ and $\phi$ spacings for all three coordinate systems, 
we usually take equidistant grid spacing.  

\begin{figure}
  \includegraphics[width=3.1in,clip]{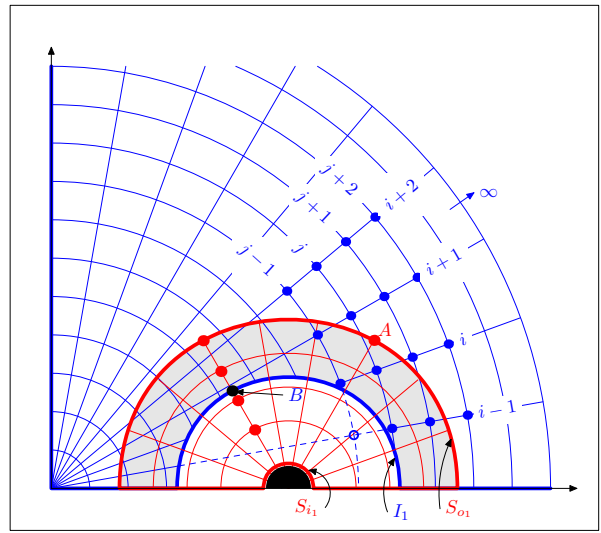}
  \caption{Schematic figure for computational domain with overlapping grids.}
  \label{fig:PE}
\end{figure}

\subsection{Finite differencing}

Standard finite difference scheme is applied to evaluate 
the derivatives of the sources 
and their numerical integrals in Eq.~(\ref{eq:GreenIde}).  
The derivatives of source terms are calculated using fourth order 
Lagrange formula, and the integrals using either trapezoidal rule 
or fourth order Simpson rule 
in $\theta$ and $\phi$ coordinates and 
second order mid-point rule for the $r$ coodinate.  
For the surface integrals at the interfaces $S_{o_{1}}$ 
and $S_{o_{2}}$ in Fig.~\ref{FigTotalCG}, the field and 
its derivatives are evaluated 
from the nearby 64 points of CCS as shown in 
Fig.~\ref{fig:PE} 
(a point A on $S_{o_{i}}$), 
to which the fourth order intepolation is applied.

\subsection{Iteration procedure}
\label{sec:iteration}

The KEH method at the $n$-th iteration follows the procedure, 
\begin{itemize}
\item[1)] Compute all the source terms in Eq.~(\ref{eq:GreenIde}).  
\item[2)] Call the Poisson Solver (described below) for each 
   of  the variables and compute their new values $\Phi^{(n)}$.  
\item[3)] Compare these newly computed values $\Phi^{(n)}$ with those 
   of previous iteration $\Phi^{(n-1)}$.  \\
   $\bullet$ If the difference is less than your accepted error 
   $\Rightarrow$  \textbf{convergence} \\
   $\bullet$ If not, update $\Phi^{(n)}$, according to 
\beq
           \Phi^{(n)}:=c\:\Phi^{(n)}+(1-c)\:\Phi^{(n-1)}, 
\label{eq:iter}
\eeq
   and go back to step (1). 
\end{itemize} 

We conclude convergence of the iteration when the difference between
two successive iterations becomes small as defined by
\beq
\frac{2|\Phi^{(n)} - \Phi^{(n-1)}|}{|\Phi^{(n)}| + |\Phi^{(n-1)}|} 
< \epsilon_{\rm c}, 
\eeq
where $\epsilon_{\rm c}=10^{-8}$ is taken in typical calculations.  
The iteration usually converges successfully, taking the convergence factor 
$c$ to be around $0.5$ when a fluid source is present.  
For the binary black hole case, it is even possible to achieve 
convergence with the factor $c=1$.  

The Poisson Solver for a potential at the $n$th iteration performs 
the following sequence of instructions :
\begin{itemize}
\item[1)] Compute the potential and its radial derivative      
   at the outer boundary of the black hole grid by      
   interpolating from nearby points of the central      
   grid.
\item[2)] Compute the surface integrals at the outer           
   boundaries of the black hole grids by using the                      
   potential and its derivative from step (1).
\item[3)] Compute the surface integrals at the inner           
   boundaries of the black hole grids by using the      
   boundary conditions for the potential or its         
   derivative.                                          
\item[4)] Compute the volume integrals inside the black        
   hole grids. Add the contributions from steps          
   (2) and (3), to obtain the value of the potential    
    inside the black hole grids.                        
\item[5)] Interpolate using points of the black hole grids     
   to compute the potential and its derivative on      
   the interfaces $I_{1}$ and $I_{2}$ of the central    
   grid.                                                
\item[6)] Compute the surface integrals on $I_{1}$ and $I_{2}$ 
   by using results from step (5).                      
\item[7)] Compute the surface integral at the outer            
   boundary of the central grid by using the            
   boundary conditions for the potential or             
   its derivative.                                      
\item[8)] Compute the volume integral inside the central       
   grid. Add the contributions from steps (6) and       
   (7), to obtain the value of the potential inside     
   the central grid.                                  
\end{itemize}

\section{Code test}
\label{sec:codetest}

In this section, we show the results for the convergence test of our 
new code.  
In the first test we compute the Newtonian potential of two spherical masses. 
Then we compute simple models for time symmetric black hole data.

As mentioned in the previous section, the local truncation errors of the
finite differencing used in our code are of order $O(\delta r^{2})$,
$O(\delta \theta^{4})$, and $O(\delta \phi^{4})$ in each coordinate. Also we have
a truncation error from multipoles higher than $L \sim 10$.  Since the local
truncation error at each grid point is a linear combination of these, and
because of the excision used in CCS, we find a non-uniform convergence 
of relative errors as we increase the resolution, as well as a non-uniform 
distribution of the errors in space, as shown below.  However, overall 
convergence is faster than second order for all cases with a fixed $L$.

\subsection{Convergence test for the Newtonian potential}

\subsubsection{Set up for the test problem}

The Poisson equation (\ref{eq:Poisson}) with a spherical source of the form
\be
S(r)=\left\{ \begin{array}{ll}
\displaystyle  \frac{(R^{2}-r^{2})^{2}}{R^{4}} & \mbox{if $0\leq r \leq R$}, \\[3mm]
               0                               & \mbox{if $r>R$},
             \end{array} \right. 
\label{TestSource}
\ee
has the solution 
\beqn
\Phi=\left\{ \begin{array}{l} 
\displaystyle  
-\frac{R^2}{6}\left[1
-\left(\frac{r}{R}\right)^{2}
+\frac35\left(\frac{r}{R}\right)^{4}
-\frac17\left(\frac{r}{R}\right)^{6}\right] 
\\[3mm] \qquad\qquad\qquad\qquad\qquad\qquad 
\mbox{if $0\leq r \leq R$}
\\[3mm]
\displaystyle  -\frac{8 R^{3}}{105\, r} 
\qquad\qquad\qquad\qquad\quad
\mbox{if $r \geq R$}
\end{array} \right.
\label{TestPotential}
\eeqn 
where $r$ is the radial coordinate and $R$ a constant.  
\begin{table}
\begin{tabular}{clrrrrrrrrrr}
\hline
Type&Coordinate&$r_a$&$r_b$&$r_c$&$d$&$N_r$&$n_r$&$n_v$&$N_\theta$&$N_\phi$&$L$ \\
\hline
S1 &CCS   & 0&  100& 3&---& 80& 40& ---& 20& 80& 10\\
   &BHCS-1& 0& 1.25& 0&1.5& 30&  0&   6& 10& 40&  5\\
\hline
S2 &CCS   & 0&  100& 3&---&160& 80& ---& 40&160& 10\\
   &BHCS-1& 0& 1.25& 0&1.5& 60&  0&  12& 20& 80&  5\\
\hline
S3 &CCS   & 0&  100& 3&---&320&160& ---& 80&320& 10\\
   &BHCS-1& 0& 1.25& 0&1.5&120&  0&  24& 40&160&  5\\
\hline
\end{tabular}
\caption{Coordinate parameters, and the number of grid points 
for each coordinate system with different resolutions.  
Each resolution is double the one above.
The parameters for BHCS-2 are identical to those of BHCS-1.
$L$ is the highest multipole included in the Legendre expansion.}
\label{tab:parameters}
\end{table}

\begin{figure}
\includegraphics[width=3.1in,clip]{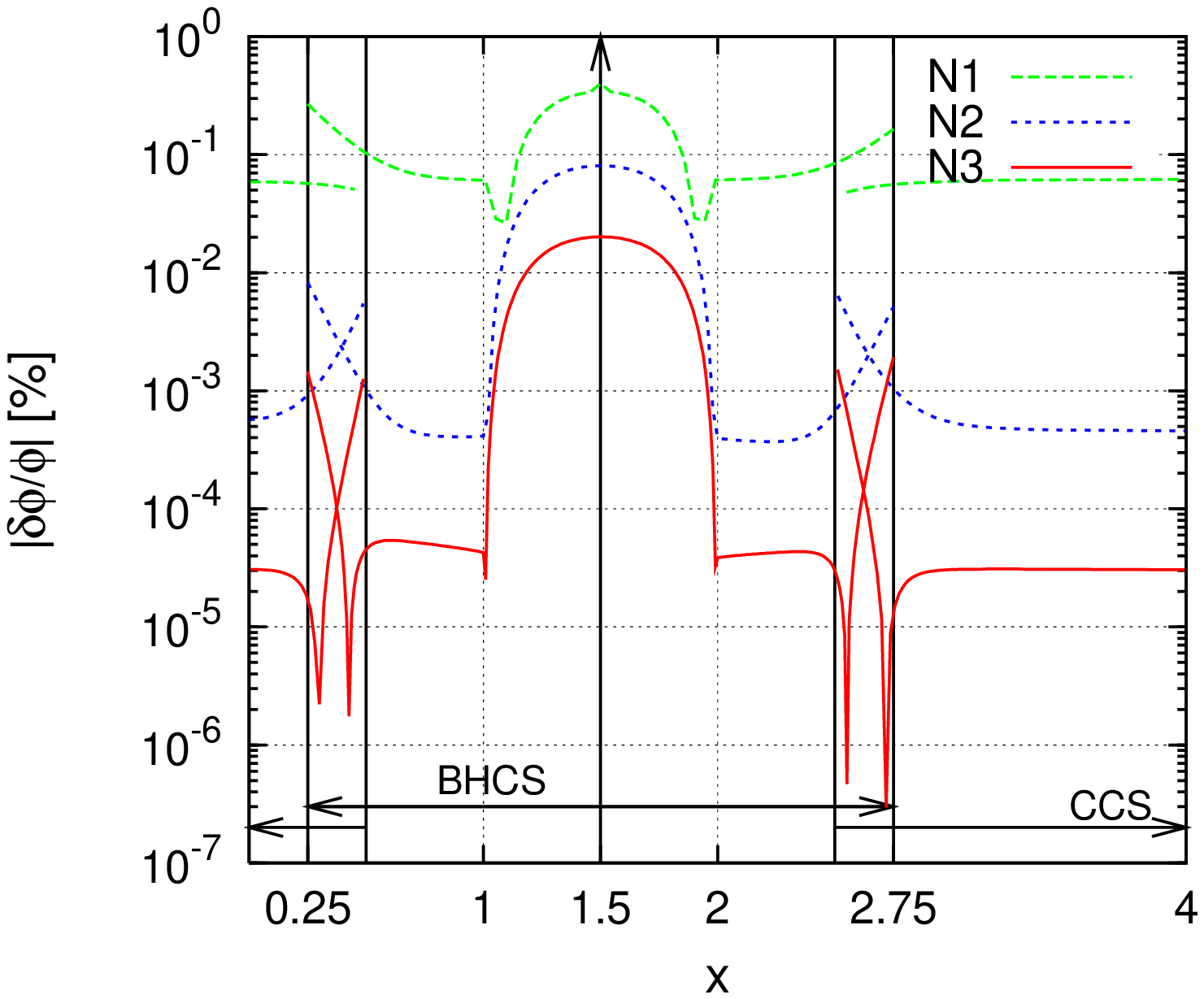}
\includegraphics[width=3.1in,clip]{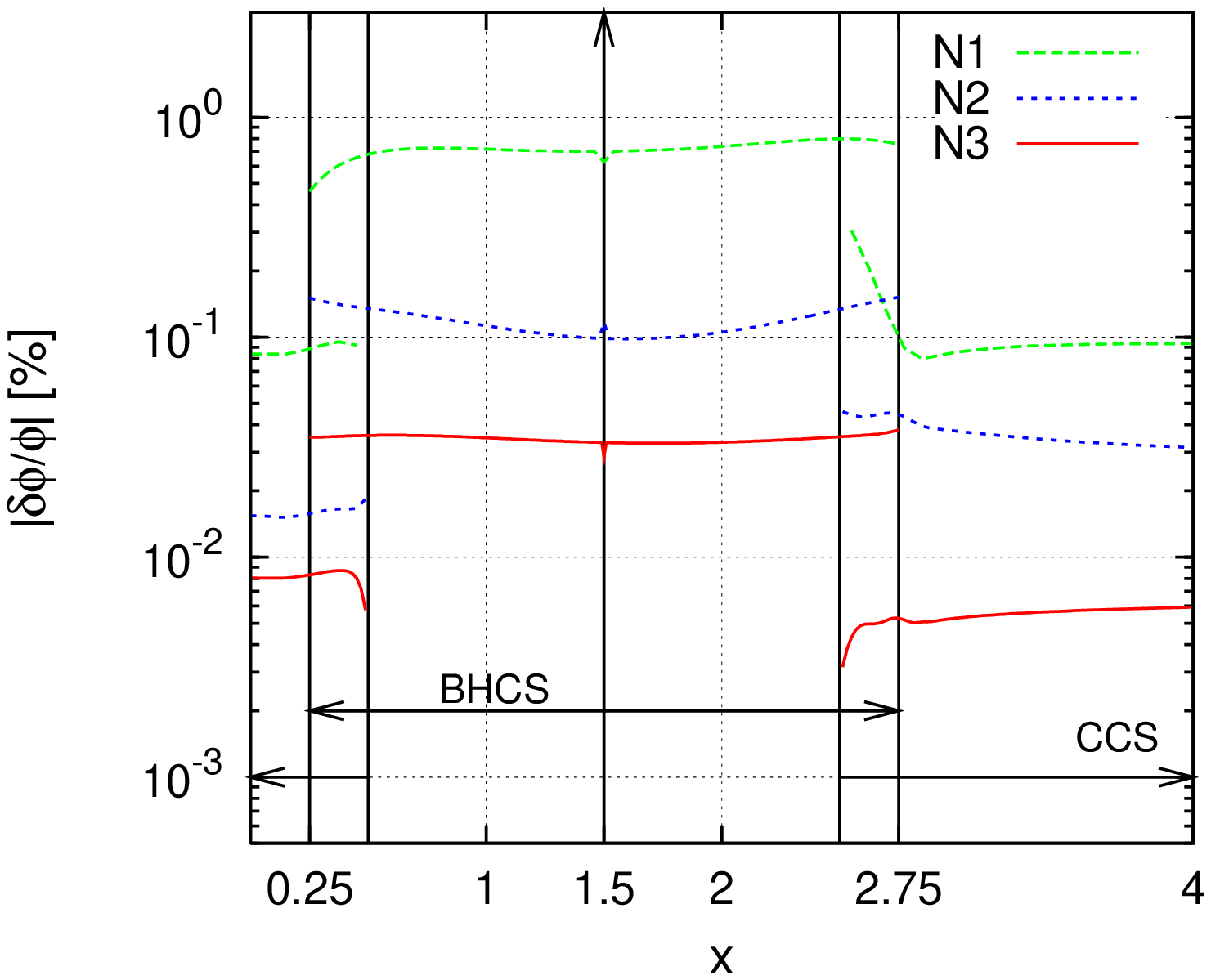}
\caption{Percentage of relative error for the source of type 
(\ref{TestSource}) for three different resolutions. Each line
from top to bottom corresponds to the resolutions S1, S2, and
S3 in table I. Vertical lines are the location of the boundary of 
numerical domains. 
Top panel: The source is inside BHCS, 
$R = 0.5 < r_b = 1.25$.
Bottom panel:
The source extends outside of the BHCS, 
$R = 1.4 > r_b = 1.25$. }
\label{TestPlot}
\end{figure}

The source (\ref{TestSource}) 
is centered at each BHCS-1 and BHCS-2, whose positions in CCS 
are $(r,\theta,\phi)=(1.5,\pi/2,0)$ and $(1.5,\pi/2,\pi)$, and 
the radii of BHCS-1 $(S_{o_{1}})$ and 2 $(S_{o_{2}})$, 
extend up to $r_b=1.25$.  The radii of the 
excised spheres in CCS $I_1$ and $I_2$ are taken as $r=1.0$.  
The radial coordinate $r$ of CCS is equidistant until $r_{c}=3$ and 
from that point until $r_b=100$ is non-equidistant, while 
the BHCSs are equidistant in the radial coordinate.  
The exact potential of two sources is a superposition of solutions 
(\ref{TestPotential}) centered at each of the two BHCS.  
For the boundary condition at $r=100$ in CCS, 
we put the potential to have its exact value.
The KEH iteration explained in previous sections is applied 
to calculate the potential $\Phi$ until convergence is made.

\subsubsection{Accuracy of numerical solutions}

In Fig.~\ref{TestPlot}, we show, for two cases, 
the percentage of the relative error between the numerical and 
exact solution, 
\beq
\left|\frac{\dl \Phi}{\Phi} \right|\, [\%]:=
100\left|\frac{\Phi_{\rm exact}-\Phi_{\rm numerical}}{\Phi_{\rm exact}}\right|.  
\label{eq:fracerror}
\eeq
In the first case (top panel), the radius of the source $R$ is 
$R=0.5$, which is smaller than the boundary radius $r_b = 1.25$ 
of BHCS-1 and BHCS-2.  For the second case (bottom panel), 
the source radius $R=1.4$ is taken so that the sources extend to CCS.  
The error is plotted along the radial coordinate 
at $(\theta,\phi)=(\pi/2,0)$ (see Fig.~\ref{FigTotalCG}).  
Along this line (labeled $x$ in the figures) BHCS-1 extends from 
$x=0.25$ to $2.75$, the source (for $R=1.4$) from $x=0.1$ to 
$2.9$, and overlap of CCS and BHCS-1 from $x=0.25$ to $0.5$ and from 
$2.5$ to $2.75$ in CCS.  The errors 
in the interval $x\in[0,4]$ are shown in the plots.  

In Fig. \ref{TestPlot}, the resolution doubles from the top to bottom 
(dashed, dotted, and solid) lines in each panel 
whose parameters are shown in Table \ref{tab:parameters}.  
The errors of the lowest resolution, top lines in each panel of
 Fig.~\ref{TestPlot}, are fairly small. 
For the case with $R=0.5$ in the top panel,
the error drops $1/4$ near the center of the source as we double the
resolution, which shows the second order convergence.
For the case of larger source in the bottom panel, 
the volume integration in CCS introduces a truncation error that 
behaves differently from the former case. Regardless of that  
the error drops again roughly as $1/4$.  

It is remarkable that the potential of such binary sources can be 
accurately computed with the small number of multipoles as $L=10$ 
in CCS.  In previous work \cite{USE00}, for binary 
neutron stars, in which only one domain corresponding to CCS 
was used to compute the field, 
a summation of more than 30 multipoles was required 
to obtain an accuracy of order 0.01\% (regarding the relative errors) that
we obtain here.

\subsubsection{Convergence property of iterations}

In test problems presented in this section, 
the number of iterations required to achieve 
convergence is about 10-20.  
Generally convergence can be accelerated using 
(1) a good initial guess for starting the iteration, and 
(2) a carefully chosen convergence factor $c$
in Eq.~(\ref{eq:iter}). \footnote{When one iterates the 
fluid variables together with the gravitational fields, 
such as a computation for binary neutron star equilibrium, the 
number of iteration may increase as many as a few hundreds.}
If the source is very steep and the convergence 
factor large (near 1.0) the iteration could blow up instead of converge. 
The number of iterations also depends critically on the size of the overlap region 
between the surfaces $S_{o_{i}}$ and $I_{i}$ $(i=1,2)$ 
(see Appendix~\ref{sec:Aptoy}).  
For the results shown in Fig.~\ref{TestPlot}, in which the overlap
is $20\%$ of the radii of $S_{o_{i}}$, 
the solution is evaluated after 14 iterations with a convergence factor $c=0.8$, 
starting from a constant (zero) potential. 
For this problem, we can achieve convergence with 
a convergence factor $c=1.0$ in 11 iterations with the same overlap region.  
On the other hand, when the overlap is set about $2.4\%$ of the radii of 
$S_{o_{i}}$, the iteration does not converge with $c=0.8$, 
and it does for $c=0.1$ after $168$ iterations.  
Although the larger overlap region is favourable for having the number 
of iteration smaller, the radii of $I_{i}$ has to be taken large enough 
to keep the number of multipoles used in CCS small.  Our choice 
for the radii of $S_{o_{i}}$ and $I_{i}$ meets these two requirements.  

We also observed that our method produces the same solution irrespective of 
the value of the convergence factor, in the above range 
$0.1 \le c \le 1$, as long as the iteration converges.

\subsection{Convergence test for solutions with excision boundaries}

\subsubsection{Analytic solutions}

To test our elliptic solver for the case with black hole excision 
boundaries, we consider the following simple solutions 
which model one or two black holes.  
For the metric 
\be
ds^{2}=-\GA^{2}dt^{2}+\GC^{4}f_{ij}dx^{i}dx^{j} \:,
\ee
where $f_{ij}$ is the flat metric, 
the hamiltonian constraint and 
the spatial trace of the Einstein equation $G_{\albe}\gamma^{\albe}=0$, 
give 
\be
\nabla^{2}\GC=0 \:\:\:\:\:\: \textrm{and} \:\:\:\:\:\: \nabla^{2}(\GA\GC)=0 \: .
\label{eq:Laplace}
\ee 
These equations have solutions, 
\be
\GC=1+\frac{M}{2r} \:\:\:\:\:\: \textrm{and} \:\:\:\:\:\: \alpha\psi=1-\frac{M}{2r}, 
\label{eq:1bhsol}
\ee
which correspond to the Schwarzschild solution with 
mass $M$, in isotropic coordinates, 
$\GC|_{r \rightarrow \infty} = 1$, and $\GA|_{r \rightarrow \infty} = 1$.  

We compute the solution (\ref{eq:1bhsol}) 
numerically by imposing boundary conditions at the sphere $r_a=M/2$.  
In order to test the code using Dirichlet boundary conditions, 
we set the boundary value at $r=r_a$ to the exact value computed from 
(\ref{eq:1bhsol}).  
For testing Neumann boundary condition, we take the value of the 
derivative of (\ref{eq:1bhsol}).  

For example, the Neumann condition for $\alpha\psi$ becomes
\beq
\frac{\pa (\alpha\psi)}{\pa r} = \frac{M}{2r^2} 
\quad \mbox{at} \quad r=\frac{M}{2}.  
\eeq
Note that the method of images, \cite{BY80}, is identical 
to requiring that $\psi$ satisfy the Robin boundary condition 
\be
\frac{\OO \GC}{\OO r}+\frac{\GC}{2r}=0  \:\:\:\:\:\: \textrm{at} \:\:\:\:\: r=M/2 \: .
\label{eq:bconpsi}
\ee
The boundary condition for the lapse, 
\beq
\alpha=0, 
\label{eq:bconalpha}
\eeq
yields the solution (\ref{eq:1bhsol}), antisymmetric about  
$ r=r_a $.  

One can also construct a two black hole solution that satisfies imaging conditions 
(\ref{eq:bconpsi}) and (\ref{eq:bconalpha}) at two spheres \cite{BY80}.  
Instead of using solutions of this kind, we use the Brill-Lindquist 
solution \cite{BL63} to Eq.~(\ref{eq:Laplace}) which is sufficient for 
the purpose of our code test.  Writing coordinates of 
BHCS-1 with subscript $1$ and BHCS-2 with subscript $2$, we write 
a two black hole solution to Eq.~(\ref{eq:Laplace}),  
\be
\GC=1+\frac{M_1}{2r_1}+\frac{M_2}{2r_2} 
\ \  \textrm{and} \ \  
\alpha\psi=1-\frac{M_1}{2r_1}-\frac{M_2}{2r_2}. 
\label{eq:2bhsol}
\ee
We set the radii of excision boundaries at $r_1=M_1/2$ and $r_2=M_2/2$, 
and impose either the Dirichlet boundary condition, which is given by 
the values of Eq.~(\ref{eq:2bhsol}) at the boundaries, 
or the Neumann boundary condition, which assigns the values of the derivatives 
of $\psi$ and $\alpha\psi$ as 
\beqn
\frac{\pa\psi}{\pa r_1} = -\frac{M_1}{2r^2_1} 
- \frac{M_2}{2r^2_2}\frac{\pa r_2}{\pa r_1} 
\ \ &\mbox{at}&\ \ 
r_1=\frac{M_1}{2}, 
\label{eq:2bhbcon_psi}
\\
\frac{\pa(\alpha\psi)}{\pa r_1} =\frac{M_1}{2r^2_1} 
+ \frac{M_2}{2r^2_2}\frac{\pa r_2}{\pa r_1} 
\ \ &\mbox{at}&\ \ 
r_1=\frac{M_1}{2}, 
\label{eq:2bhbcon_alp}
\eeqn
and $1 \leftrightarrow 2$ for the boundary at $r_2=M_2/2$.  
The coordinates $r_1$ and $r_2$ are written in terms of each other as 
\beqn
&& r_1 = \sqrt{r_2^2 - 2 r_2 a \sin\theta_2\cos\phi_2 + a^2},
\\
&& r_2 = \sqrt{r_1^2 + 2 r_1 a \sin\theta_1\cos\phi_1 + a^2}.
\eeqn
For each boundary condition, the accuracy of our Poisson solver 
is examined comparing these solutions to the analytic ones.  
\footnote{In the above two black hole solution, the lapse $\alpha$
takes negative value in the neighborhood of each boundary sphere.  
}

The Laplace equations (\ref{eq:Laplace}) are solved from 
the surface integrals at the boundaries in our Poisson solver.   
The same equations (\ref{eq:Laplace}) can be rewritten as
\be
\nabla^{2}\GC=0 \:\:\:\:\:\: \textrm{and} \:\:\:\:\:\: 
\nabla^{2}\GA=-\frac{2}{\GC}f^{ij}\OOD{i}\GC \OOD{j}\GA .
\label{eq:Laplace2}
\ee 
This form is 
also used to test the volume integral over the source 
in the Poisson solver.  

In the next two sections $M_1/2$, $M_2/2$ refer to the radii of 
$S_{i_1}$ and $S_{i_2}$ of BHCS-1 and BHCS-2 correspondingly.

\subsubsection{Convergence test for one black hole solution}

\begin{figure}
  \includegraphics[width=3.4in,clip]{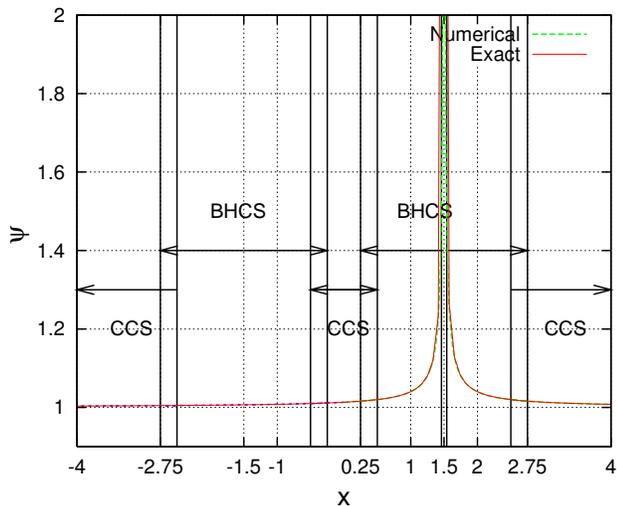}
  \caption{Exact and numerical solution for $\GC$ on the x-axis 
  using the boundary condition (\ref{eq:bconpsi})}.
  \label{FigLapse}
\end{figure}

First we treat the problem with no volume sources as in equations
(\ref{eq:Laplace}). In Fig. \ref{FigLapse}, 
the conformal factor $\psi$ is plotted along the x-axis.  
%
BHCS-1 and 2 are centered on the x-axis at $x=1.5$ and $-1.5$ 
respectively.  
In BHCS-1, we have two surface integrals, one at the 
inner (excision) boundary, $S_{i_{1}}$, with a radius $r_a=0.02$ 
(in conformal geometry) and one at the outer boundary, $S_{o_{1}}$, 
at a distance $r_b=1.25$.  
%
In the BHCS-2, there is no inner boundary sphere $S_{i_{2}}$; 
we solve for the whole region inside $S_{o_{2}}$ without any excised
region.  
In this grid we need only to compute one surface integral at $S_{o_{2}}$. 
Finally CCS extends to a distance $r=100$ and it excludes regions inside 
of two spheres
$I_{1}$ and $I_{2}$ which are centered at $x=1.5$ and $x=-1.5$
respectively and have radius $1.0$. The overlapping region is one shell 
centered at $x=1.5$ with $1.0\leq r \leq 1.25$ and the corresponding
one on the negative x-axis. 

\begin{figure}
  \includegraphics[width=3.4in,clip]{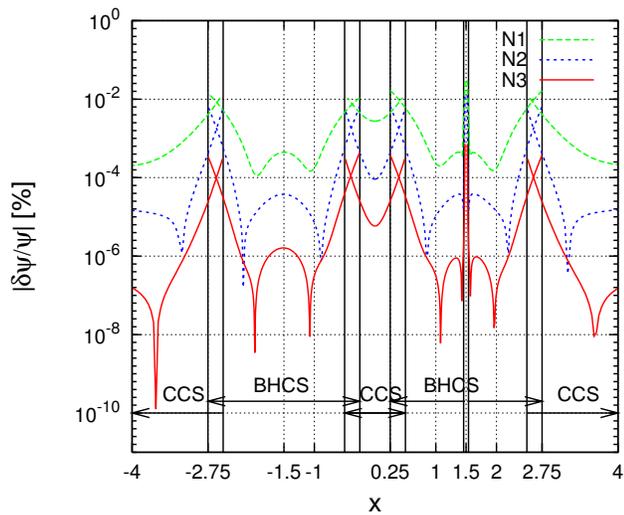}
  \caption{Fractional errors of the conformal factor $\psi$,  
are plotted along the x-axis. Lines from top to bottom corresponds 
to the resolutions N1, N2, and N3 in Table \ref{tab:param_onethroat_nosource}.  
Vertical lines are the boundaries of numerical domains.}
  \label{FigLapseFE}
\end{figure}

In Fig.~\ref{FigLapseFE}, we show the fractional errors of the conformal factor 
shown in Fig.~\ref{FigLapse} for three different resolutions 
in table \ref{tab:param_onethroat_nosource}.  The error of $\GA$ is of the same order. 
Since the terms calculated using 2nd order finite differencing, 
such as volume integrals, do not contribute in this solution, 
4th-order convergence can be seen in Fig.~\ref{FigLapseFE} as expected.

\begin{table}
\begin{tabular}{clrrrrrrrrrr}
\hline
\multicolumn{12}{c}{One black hole data without source} \\
\hline
Type&Coordinate &$r_a$&$r_b$&$r_c$&$d$&$N_r$&$n_r$&$n_v$&$N_\theta$&$N_\phi$&L \\
\hline
N1 &CCS   & 0   &  100&   3&---& 80& 40& ---& 20& 80& 10\\
   &BHCS-1& 0.02& 1.25&0.02&1.5& 30&  0&   6& 10& 40&  5\\
   &BHCS-2&    0& 1.25&   0&1.5& 30&  0&   6& 10& 40& 5\\
\hline
N2 &CCS   & 0   &  100&   3&---&160& 80& ---& 40&160& 10\\
   &BHCS-1& 0.02& 1.25&0.02&1.5& 60&  0&  12& 20& 80& 10\\
   &BHCS-2&    0& 1.25&   0&1.5& 60&  0&  12& 20& 80& 10\\
\hline
N3 &CCS   & 0   &  100&   3&---&320&160& ---& 80&320& 10\\
   &BHCS-1& 0.02& 1.25&0.02&1.5&120&  0&  24& 40&160& 10\\
   &BHCS-2&    0& 1.25&   0&1.5&120&  0&  24& 40&160& 10\\
\hline
\end{tabular}
\caption{Grid parameters used in one BH test problem.  
The same conventions as in Table \ref{tab:parameters} are used.}
\label{tab:param_onethroat_nosource}
\end{table}

\begin{figure}
  \includegraphics[width=3.4in,clip]{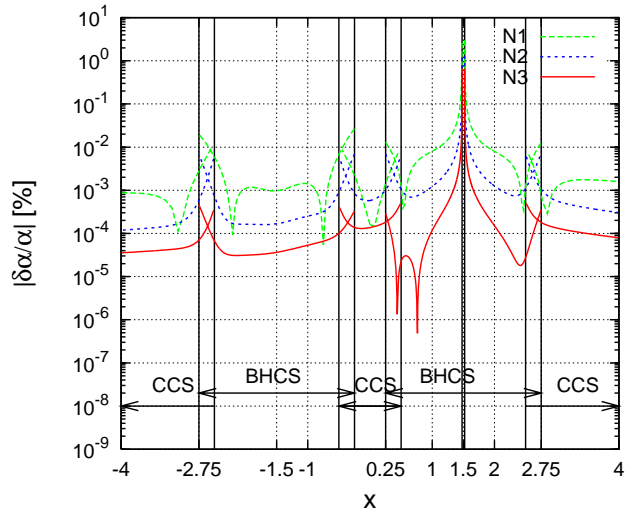}
  \caption{Fractional errors of the lapse $\alpha$,  
are plotted along the x-axis. Lines from top to bottom corresponds 
to the resolutions N1, N2, and N3 in Table \ref{tab:param_onethroat_source}.  
The no-boundary Green's function $G^{\rm NB}$ has been used.}
  \label{FigLapse_G_S_OneThroat}
\end{figure}

\begin{figure}
  \includegraphics[width=3.4in,clip]{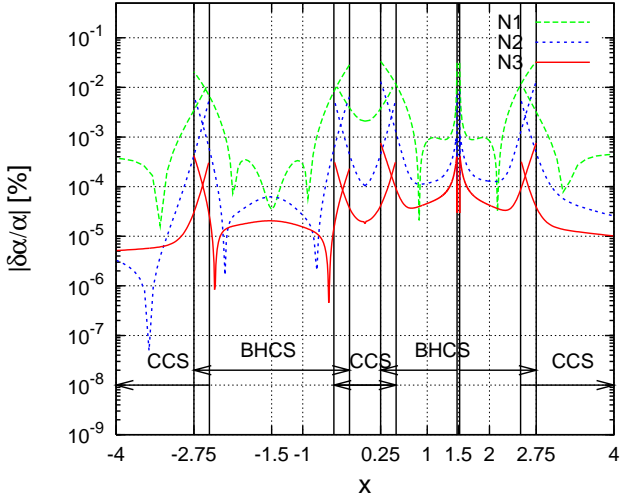}
  \caption{Same as in Fig. \ref{FigLapse_G_S_OneThroat}, but now $G^{\rm DD}$ is
used.}
  \label{FigLapse_Gdd_S_OneThroat}
\end{figure}

Next we solve the same problem but in the form of (\ref{eq:Laplace2}).  Near the inner 
boundary of BHCS-1, the source for the volume integral of the lapse, becomes very steep
(approximately  it goes as $r^{-4}$) and needs more grid points than the previous
case in order to achive the same order of error.  For a grid set up as the one
shown in Table \ref{tab:param_onethroat_source}, the fractional errors of the lapse are shown in 
Fig. \ref{FigLapse_G_S_OneThroat} where the no-boundary Green's function 
$G^{\rm NB}$ has been used. The error for the conformal factor is as in Fig. \ref{FigLapseFE} since again we 
don't have any volume sources to integrate.
For the same problem if we use the $G^{\rm DD}$ Green's function we get
the fractional errors of Fig. \ref{FigLapse_Gdd_S_OneThroat}. With this latter choice
of the Green's function the error near the throat is much smaller. Also we need fewer iterations to achieve convergence. 

\begin{table}
\begin{tabular}{clrrrrrrrrrr}
\hline
\multicolumn{12}{c}{One black hole data with source} \\
\hline
Type&Coordinate &$r_a$&$r_b$&$r_c$&$d$&$N_r$&$n_r$&$n_v$&$N_\theta$&$N_\phi$&L \\
\hline
M1 &CCS   & 0   &  100&  3&---& 80& 40& ---& 20& 80& 10\\
   &BHCS-1& 0.02& 1.25&1.0&1.5&120&112&   8& 10& 40&  5\\
   &BHCS-2&    0& 1.25&  0&1.5& 30&  0&   6& 10& 40&  5\\
\hline
M2 &CCS   & 0   &  100&  3&---&160& 80& ---& 40&160& 10\\
   &BHCS-1& 0.02& 1.25&1.0&1.5&240&224&  16& 20& 80& 10\\
   &BHCS-2&    0& 1.25&  0&1.5& 60&  0&  12& 20& 80& 10\\
\hline
M3 &CCS   & 0   &  100&  3&---&320&160& ---& 80&320& 10\\
   &BHCS-1& 0.02& 1.25&1.0&1.5&480&448&  32& 40&160& 10\\
   &BHCS-2&    0& 1.25&  0&1.5&120&  0&  24& 40&160& 10\\
\hline
\end{tabular}
\caption{Grid parameters used in one BH test problem with source.  
The same conventions as in Table \ref{tab:parameters} are used.}
\label{tab:param_onethroat_source}
\end{table}

\subsubsection{Convergence test for binary black hole solution}

\begin{figure}
\includegraphics[width=3.4in,clip]{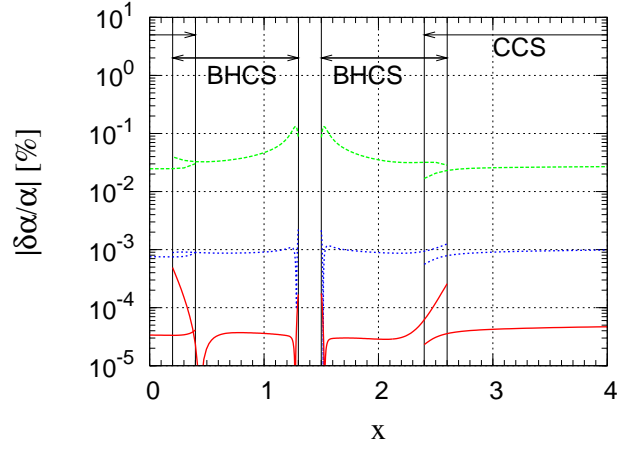}
\caption{
Fractional error of the lapse 
is plotted along the x-axis.  
Lines from top to bottom correspond 
to the resolutions T1, T2, and T3 in Table \ref{tab:param_toybh}.  
A set of Eq.~(\ref{eq:Laplace}) is solved using the no-boundary 
Green's function and by imposing Neumann boundary conditions.  
Vertical lines correspond to the boundaries of the numerical domains.
}
\label{fig:bbherror1}
\end{figure}

The two black hole solution (\ref{eq:2bhsol}) is computed numerically solving 
either set of (\ref{eq:Laplace}) or (\ref{eq:Laplace2}).  
In Fig.~\ref{fig:bbherror1}, 
we plot the fractional error, Eq. \ref{eq:fracerror} for the lapse $\alpha$
computed from the first set of Eq.~(\ref{eq:Laplace})
using the no-boundary Green's function $G^{\rm NB}$ 
and Neumann boundary condition.  

\begin{table}
\begin{tabular}{clrrrrrrrrrr}
\hline
\multicolumn{12}{c}{Two black hole data} \\
\hline
Type&Coordinate &$r_a$&$r_b$&$r_c$&$d$&$N_r$&$n_r$&$n_v$&$N_\theta$&$N_\phi$&L \\
\hline
T1 &CCS   & 0  & 100& 2.8& ---& 52& 28& ---&16& 32&   6\\
   &BHCS-1& 0.1& 1.2& 1.0& 1.4& 32& 30&   2&16& 32&  10\\
\hline
T2 &CCS   & 0  & 100& 2.8& ---&104& 56& ---&32& 64&  10\\
   &BHCS-1& 0.1& 1.2& 1.0& 1.4& 64& 60&   4&32& 64&  10\\
\hline
T3 &CCS   & 0  & 100& 2.8& ---&208&112& ---&64&128&  10\\
   &BHCS-1& 0.1& 1.2& 1.0& 1.4&128&120&   8&64&128&  10\\
\hline
\end{tabular}
\caption{
Grid parameters used in two BH test problem.  
The same conventions as in Table \ref{tab:parameters} are used.}
\label{tab:param_toybh}
\end{table}

The integral form of the first set (\ref{eq:Laplace}) involves 
solely the surface integrals of Eq.~(\ref{eq:GreenIde}).  
Since the surface terms are calculated using 4th order finite
differences, convergence of this order can be seen in
Fig.~\ref{fig:bbherror1} when
the grid resolution is increased from T1 to T3 in Table
\ref{tab:param_toybh}.
Starting from $\alpha=\psi=1$, the solution converges after 
$15$ iterations with the convergence factor $c=1.0$.  
Typical CPU time and memory to compute the $1/4$ of the whole 
binary black hole region is tabulated in Table \ref{tab:CPU}.

\begin{figure}
\includegraphics[width=3.4in,clip]{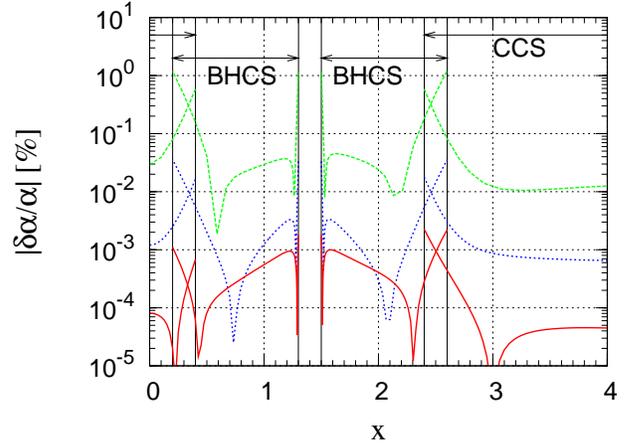}
\includegraphics[width=3.4in,clip]{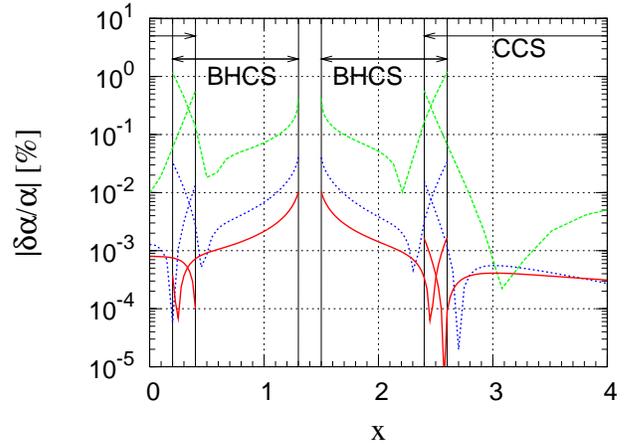}
\caption{
Same as Fig.~\ref{fig:bbherror1}, but for Eq.~(\ref{eq:Laplace2}). For the
BHCS, $G^{\rm DD}$ is used in the top panel and $G^{\rm ND}$ in the bottom panel.}
\label{fig:bbherror2}
\end{figure}

\begin{figure}
\includegraphics[width=3.4in,clip]{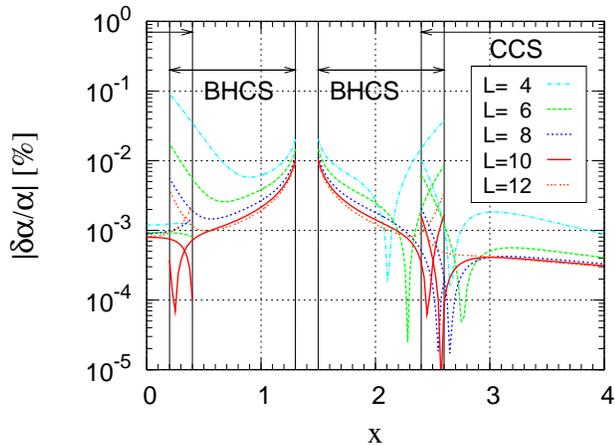}
\caption{
Fractional errors of the lapse $\alpha$ are plotted along 
the x-axis for different values of the highest multipole in the Green's function.  
A set of Eq.~(\ref{eq:Laplace2}) is solved using 
the Green's function $G^{\rm ND}$ with resolution 
T3 in Table \ref{tab:param_toybh}.  
}
\label{fig:bbherrorLtest}
\end{figure}

\begin{table}
\begin{tabular}{ccc}
\hline
Type& CPU time/iteration [s] & Memory [MB]\\
\hline
T1 &  $0.14$ &  $25$ \\
T2 &  $1.0$  &  $73$ \\
T3 &  $8.6$  & $236$ \\
\hline
\end{tabular}
\caption{
Typical CPU time and memory used for the BH
calculations.  The use of equatorial 
and $\pi$ rotation symmetries reduce 
the number of grid points indicated in Table 
\ref{tab:param_toybh} by a factor of 4.  
Note that the computational costs approximately scale 
linearly with respect to the total number of grid points 
which is $2^3=8$ times at each level T1-T3.  
Opteron 2 GHz with Portland fortran compiler is used.}
\label{tab:CPU}
\end{table}

In Fig.~\ref{fig:bbherror2}, the fractional error 
(\ref{eq:fracerror}) of the solution to Eq. (\ref{eq:Laplace2}) 
is shown for a different choice of the Green's functions for BHCS.  
The numerical integration in radial direction, that appears in the volume integrals 
of (\ref{eq:GreenIde}) are calculated with 2nd order accurate mid-point rule.  
In all test problems, we found that the largest error appears in computing 
$\alpha$ with Neumann boundary condition shown in Fig.\ref{fig:bbherror2} bottom panel.
However, even for this case, 
the error is controlled to give a fractional error less than $0.01 \%$ 
everywhere, when the highest resolution T3 is used.  
We have tested different combinations of the Green's functions with
boundary conditions, and found similar or better convergence results. 
(The fractional error of the conformal factor $\psi$, on the other hand, 
scales in 4th order since the equation for $\psi$ does not involve 
the volume integrals.  )

Finally, the convergence test for the number of multipoles 
summed in the Green's function, $L$, with a fixed resolution (T3 in Table 
\ref{tab:param_toybh}), is shown in Fig.\ref{fig:bbherrorLtest}.  
Changing $L$ from 4 to 12, convergence is achieved around $L=10$.  
This is a dramatically small number compared to the binary neutron star 
case \cite{USE00} , for which $L=30\sim 40$ multipoles are required 
because only one domain (corresponding to CCS here) was used for computing 
the gravitational fields.

\section{An example for binary black hole initial data}
\label{sec:inidata}

To conclude the test for our new numerical code, 
we show an example of binary black hole initial data; 
a binary black hole solution of IWM formulation.  
This is to demonstrate that our new code 
can produce the binary black hole initial data 
with nonzero angular momentum, and locate 
the apparent horizon using the method described in 
Appendix \ref{sec:aphzn} at the same time.  
Writing the spatially conformally flat metric 
\beq
ds^2 \,=\, -\alpha^2dt^2 + \psi^4 f_{ij} (dx^i+\beta^i dt) (dx^j+\beta^j dt),
\eeq
in a chart $\{t,x^i\}$, the 5 metric potentials, the conformal factor $\psi$, 
the shift $\beta^a$ and the lapse $\alpha$ are solved from the Hamiltonian 
constraint, momentum constraint and the spatial trace of the Einstein's 
equation, respectively.  
As shown in Appendix \ref{sec:eqs}, all these 
eqations are written in elliptic form.

At the black hole excision boundary, 
certain Dirichlet data is imposed to ensure that the apparent 
horizon appears outside of the boundary sphere. For simplicity, we
do not intend to impose 
certain physically motivated boundary conditions (see below).  
Dirichlet boundary conditions are given 
to all variables $\{ \psi, \alpha, \beta^a \}$ as 
\beqn
\psi &=& \psi_{\rm B} = {\rm constant}, 
\\
\alpha  &=& \alpha_{\rm B} = {\rm constant},   
\\
\beta^a  &=& - \Omega\,\phi_{\rm C}^a - \Omega_{\rm B}\,\phi_{\rm B}^a, 
\eeqn
at the excision sphere $r=r_a$ of BHCS. 
For the boundary value of the conformal factor $\psi_{\rm B}$, a constant is chosen 
large enough to form apparent horizons near the excision spheres.  
For the lapse $\alpha_{\rm B}$, we also assign a constant value.  
The boundary condition for the shift vector assigns a momentum and a spin 
to each hole.  Here, $\phi_{\rm C}^a$ and $\phi_{\rm B}^a$ are the basis of 
$\phi$ coordinate of CCS and BHCS, respectively.  

\begin{table}
\begin{tabular}{cccccc}
\hline
Type & $r_a$ & $\psi_{\rm B}$  & $\alpha_{\rm B}$ & 
$\Omega$ & $\Omega_{\rm B}$ \\
\hline
B1 & $0.1$ & $3.0$ & $1.0$ & $0.3$ &  $0.0$ \\
\hline
\end{tabular}
\caption{
Parameters for the boudnary conditions.  
Except for the value of $r_a$, the parameter set of T3 
in Table \ref{tab:param_toybh} is used for the computation.  
}
\label{tab:bcon}
\end{table}

\begin{figure}
\includegraphics[scale=0.7,clip]{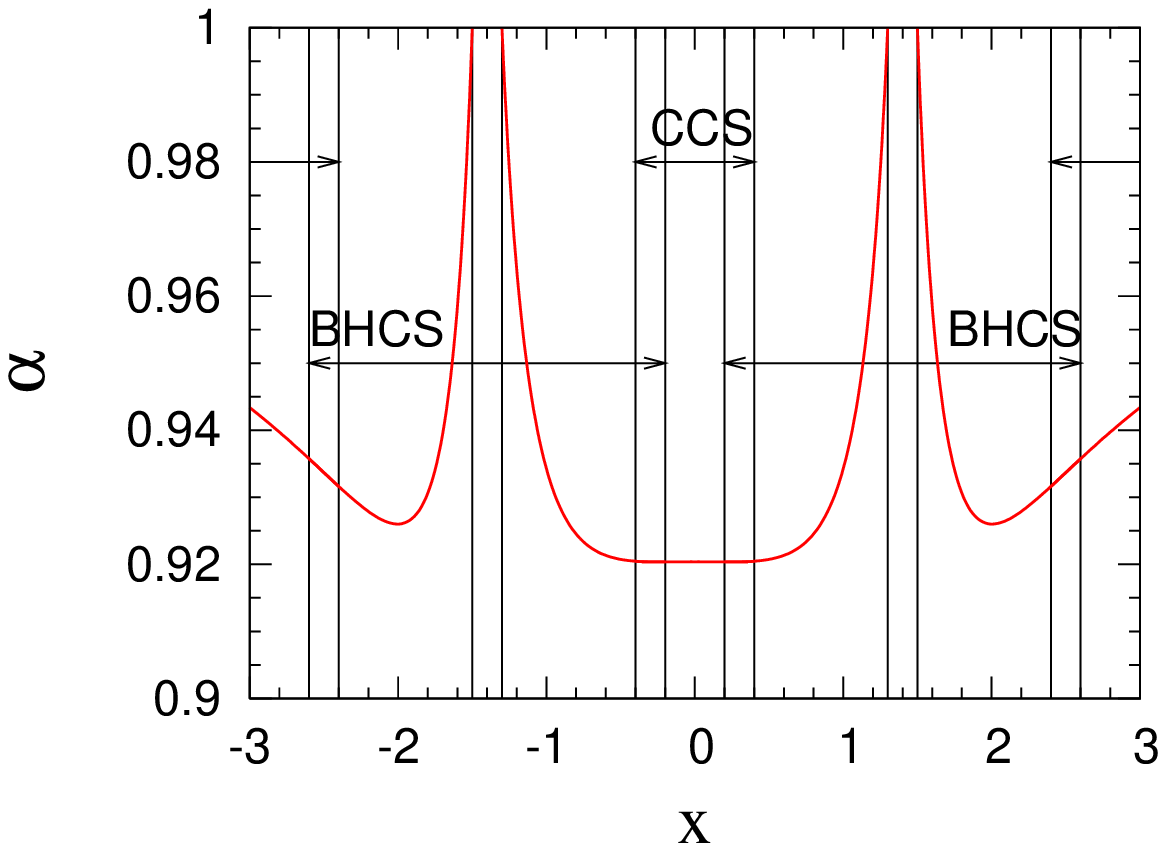}
\includegraphics[scale=0.7,clip]{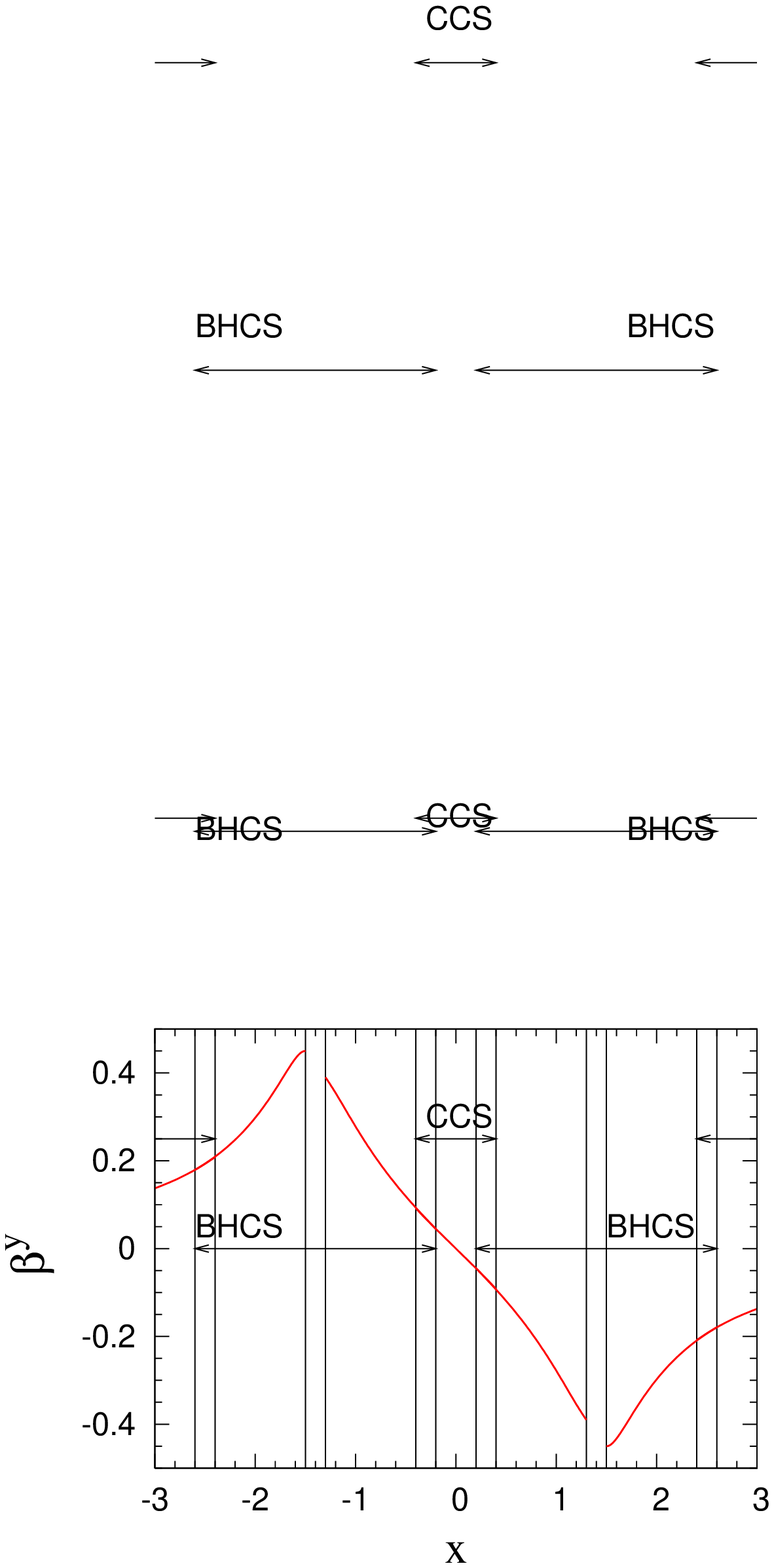}
\caption{
%
Plots for $\alpha$ (top) and $\beta^y$ (bottom) 
along the x-axis which pass through the excised region of the two holes.  
Vertical lines are the boundaries of each computational domain.  
}
\label{fig:plot_cfbbh}
\end{figure}
\begin{figure}
\includegraphics[scale=0.8,clip]{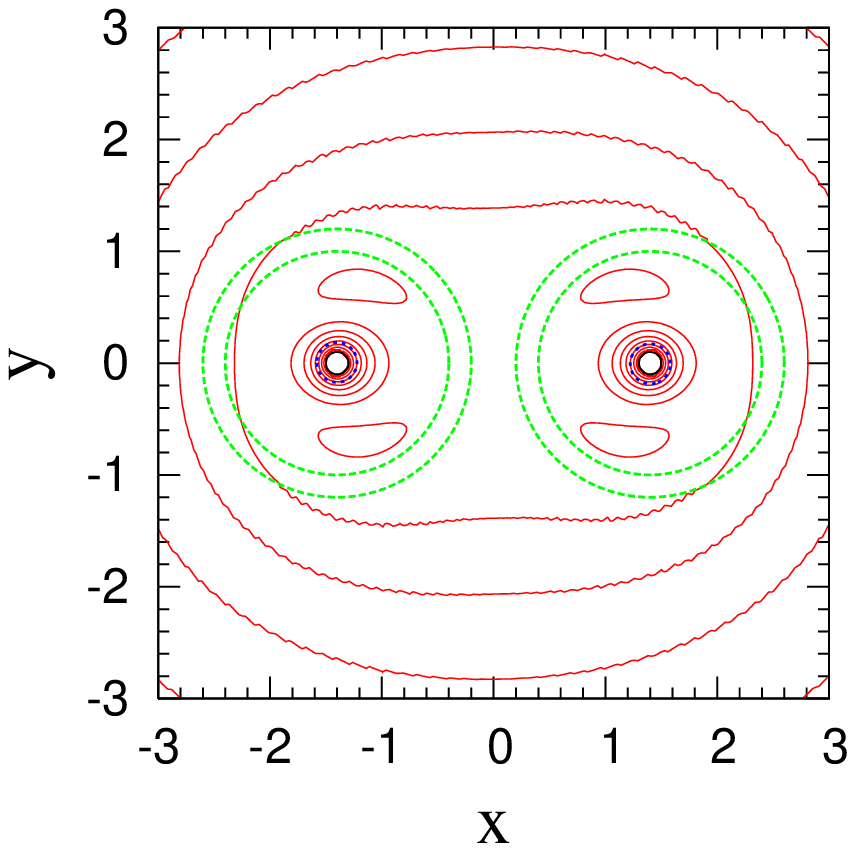}
\includegraphics[scale=0.8,clip]{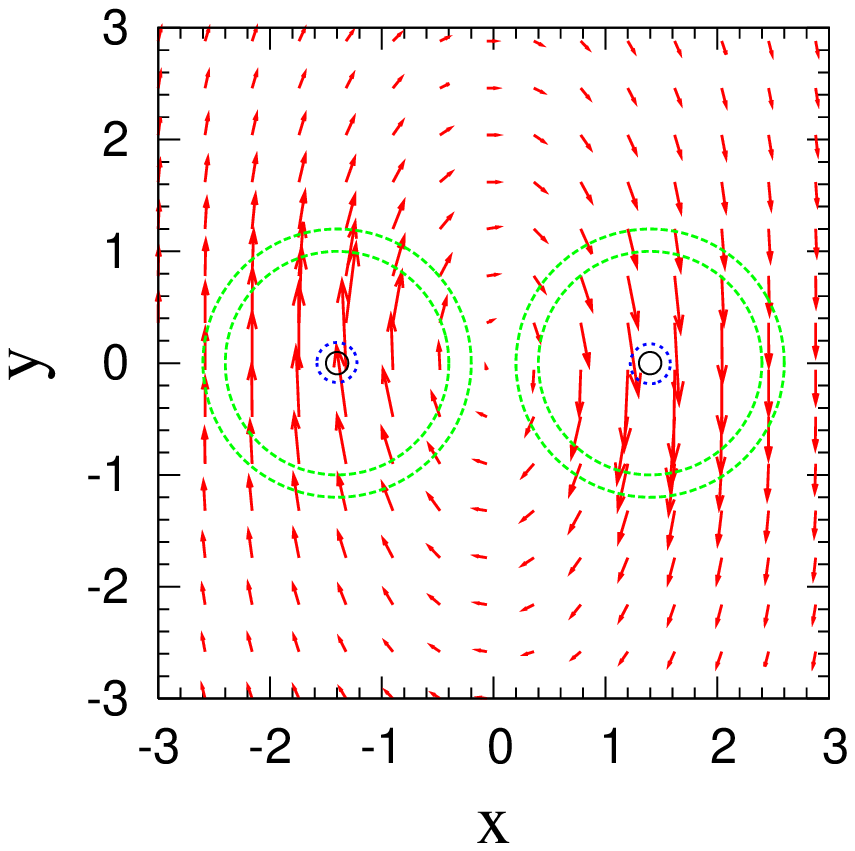}
\caption{
Contours of the lapse $\alpha$ (top) and 
the vector field of the shift $\beta^a$ (bottom) 
for the same model as Fig.\ref{fig:plot_cfbbh}.  
Thick dotted circle in each figure is the apparent 
horizon.  
}
\label{fig:cont_cfbbh}
\end{figure}

In Figs.~\ref{fig:plot_cfbbh} and \ref{fig:cont_cfbbh}, we show a solution 
with the boundary parameters shown in Table \ref{tab:bcon}.  
All potentials are smoothly joined across the overlap of CCS and BHCS.  
In Fig.\ref{fig:cont_cfbbh}, thick dotted circles right outside of 
the excised sphere (thin black circles) are the apparent horizons located 
using the method in the Appendix \ref{sec:aphzn}.

\section{Discussion}

The KEH iteration using the Green's formula Eq.~(\ref{eq:GreenIde}) 
is applicable to solve various types of partial differential equations with nonlinear sources.  
For example, this method has been applied to solve the helically symmetric 
scalar field and binary neutron stars \cite{Yoshida06}.  In this work, 
the equation for the scalar field is written in the form of Helmholtz 
equation, and the half-advanced + half-retarded Green's function is used in 
Eq.~(\ref{eq:GreenIde}) to compute a standing wave solution iteratively.  
We plan to compute the helically symmetric binary black hole/neutron 
star solution using the coordinate systems and iteration scheme presented 
in this paper.  

Black hole singularities on an initial hypersurface 
are avoided either by excising a numerical domain in the neighbourhood of the singularities, 
or by using punctures.  
In the previous Sec.~\ref{sec:inidata}, we applied rather crude 
boundary conditions at the black hole excision sphere.  
For physically motivated excision conditions, one imposes apparent horizon 
or isolated horizon boundary condition to the conformal factor at the excised sphere 
so as these spheres become automatically horizons \cite{pepe,Cook2004}.  
Alternatively one can use the puncture method to produce accurate initial data
as the ones used in binary black holes evolutions \cite{NASA2006,BROWN2006,Diener2006,HSL2006,Brug2006},
and black hole/neutron star binary simulations \cite{SU06}.

\acknowledgments

It is a pleasure to thank Spiros Cotsakis and Masaru Shibata for helpful discussions. 
Especially we would like to thank John L. Friedman for warm and continuous encouragement, 
and for providing the key idea of overlapping grids 
together with the argument of Appendix \ref{sec:Aptoy}.
This work was was supported by NSF grants PHY0071044, PHY0503366, as well as
the joint Greek Ministry of Education
and European Union research grant "Pythagoras" No.~1351.

\textit{Note added in proof}.--Independently to our work, a new apparent horizon solver
was developed recently using the same formulation by Lap-Ming Lin and Jerome Novak \cite{Lin}.

\appendix

\section{Apparent horizon solver}
\label{sec:aphzn}

A method for locating the apparent horizon is described in this Appendix.  
Our method is a modification of \cite{NKO1984} and it 
will be one of the simplest apparent horizon finders without 
any symmetry restriction.  
For other works on apparent horizon finders, 
see e.g. \cite{Thorn2004,apshi} and references therein. 

An apparent horizon ${\cal A}$ is defined as the boundary of all trapped 
regions on a spacelike hypersurface $\Sigma$, where the expansion $\vartheta$ 
of the outgoing null congruence $\ell^\alpha$ orthogonal to ${\cal A}$ vanishes.  
Introducing a null foliation ${\cal H}_u$ whose normal is $\ell^\alpha$, 
where $u$ labels a family of null hypersurfaces, the apparent horizon ${\cal A}$ 
is the intersection ${\cal A} = {\cal H}_0 \cap \Sigma$.  
To each two dimensional surface ${\cal H}_u \cap \Sigma$, 
the spatial unit normal vector $s^a$ is associated, and the outgoing 
null vector field $\ell^\alpha$ can be written with a function $f$, 
$\ell^\alpha = f(n^\alpha + s^\alpha)$, where $n^\alpha$ is a timelike 
normal to $\Sigma$.  
Then, writing a projection tensor onto a surface orthogonal to $s^a$
as $e_{ab} = \gmabd - s_a s_b$, 
the expansion $\vartheta$ is written 
\beq
\vartheta = e^{\albe}\na_\alpha \ell_\beta = f (D_a s^a - e^\albe K_\albe)
\eeq
Hence, the equation 
\beq
D_a s^a - e^{ab}K_{ab} = 0, 
\label{eq:aheq1}
\eeq
is satisfied at the apparent horizon $\cal A$.  

We introduce a family of level surfaces parametrized by 
a spherical coordinate surrounding a black hole 
in the form , 
\beq
F:= r-R_h(\theta,\phi),
\label{eq:lvsf}
\eeq  
where $F=0$ coincides with the apparent horizon.
\footnote{The level surface $F$ may not coincide with the 
intersection ${\cal H}_u \cap \Sigma$, except at $\cal A$}
The spatial normal $s^a$ is proportional to the gradient of $F$, 
\beq
s_a = \frac{D_a F}{||DF||}, 
\eeq
where the norm $||DF||$ is defined by 
\beq
||DF|| = \sqrt{\gmabu D_a FD_b F}
\label{eq:norm1}
\eeq
We rewrite Eq.~(\ref{eq:aheq1}) as an elliptic equation for the level 
surfaces in the conformally related geometry.  Introducing quantities 
weighted with the conformal factor, 
$e_{ab} := \psi^{4}\tilde e_{ab}$, 
$K_{ab} := \psi^{4}\tilde K_{ab}$, 
$s_a := \psi^{2}\tilde s_a$, the norm (\ref{eq:norm1}) is transformed 
\beq
||DF|| = \psi^{-2} \sqrt{\tgmabu \tD_a F\tD_b F} = \psi^{-2}||\tD F||
\eeq
and 
\beq
\tilde s_a = \frac{\tD_a F}{||\tD F||}. 
\eeq
Multiplying Eq.~(\ref{eq:aheq1}) by the factor $\psi^{2}||\tD F||$, 
we obtain an elliptic equation for the apparent horizon, namely 
\beqn
&&\tilde \Delta F+ \hat S =0, 
\\
&&\hat S:=\tD_a \ln\frac{\psi^4}{||\tD F||}\tD^a F
-\psi^{2}||\tD F||\tilde K_{ab} \tilde  e^{ab}, 
\eeqn
which is satisfied on the surface $r=R_h(\theta,\phi)$.  
Separating the Laplacian associated with the flat metric, $\zLap$, 
from $\tilde \Delta$ associated with the conformal metric,  
$\tgmabd$, we have
\beq
\tLap F = \zLap F 
+ h^{ab}\zD_a\zD_b F - \tgmabu C^c_{ab}\zD_c F, 
\eeq
where the second and third terms have been moved to the right hand side.  
Defining 
\beqn
\zLap F &=& \frac2{r}-\frac1{r^2}\zLap_H R_h 
= -\frac1{R_h^2}(\zLap_H -2)R_h, 
\\
\zLap_H
&:=&\frac1{\sin\theta}\frac{\pa}{\pa \theta}
\left(\sin\theta\frac{\pa}{\pa \theta}\right)
+\frac1{\sin^2\theta}\frac{\pa^2}{\pa \phi^2}, 
\eeqn
the equation for the apparent horizon (\ref{eq:aheq1}) is rewritten 
\beqn
&&(\zLap_H -2)R_h \,=\, S,
\label{eq:aheq}
\\
&&S\,:=\, R_h^2 (\tilde S + \hat S),
\\
&&\tilde S \,:=\, -h^{ab}\zD_a\zD_b F+ \tgmabu C^c_{ab}\zD_c F.
\eeqn
Terms in $\tilde S$ vanish for spatially 
conformal flat geometry.

We find a solution to Eq.~(\ref{eq:aheq}) 
\beq
R_h(x) = -\frac1{4\pi}\int_H d^2x\, G(x,x') S(x'), 
\label{eq:ahsol}
\eeq
where the coordinates $x$ here represents $(\theta,\phi)$, and the 
function $G(x,x')$ is given in terms of Legendre expansion, 
\beqn
G(x,x')
&=&\sum_{\ell=0}^{\infty}
\frac{2\ell+1}{\ell(\ell+1)+2}
\sum_{m=0}^{\ell}\epsilon_m \frac{(\ell-m)!}{(\ell+m)!}
\nonumber \\
&&\!\!\!\!\!\!\!\!\!\!\!\!\!\!\!
\times
P_\ell^m(\cos\theta)
P_\ell^m(\cos\theta')
\cos m(\phi-\phi'). 
\label{eq:ahgreen}
\eeqn
The same discretization as in BHCS, and 
an iteration similar to KEH method are applied to 
Eq.~(\ref{eq:ahsol}) (see Sec.\ref{sec:coding}).  
The 4th order Lagrange formula is used for finite differencing 
the source and for numerical integration.  The iteration 
converges typically in 30 iterations, whose CPU time is 
negligible in the computation of initial data.

\section{Multipole expansion of the Green's functions of the Laplacian.}
\label{sec:ApGreen}

In our Poisson solver (\ref{eq:GreenIde}) one must choose appropriate 
Green's functions to meet the boundary conditions imposed on each 
of the field variables.  
In this Appendix, we present explicit forms of those used in the 
preceding sections.  They are the Green's function without boundaries, 
$G^{\rm NB}(x,x')$, 
and two Green's functions with boundaries on two concentric spheres 
$S_a$ and $S_b$ at radius $r=r_a$ and $r=r_b$, where $r_a < r_b$; 
one of them imposes Dirichlet conditions on both $S_a$ and $S_b$, 
$G^{\rm DD}(x,x')$, and the other imposes Neumann condition on 
$S_a$ and Dirichlet condition on $S_b$, $G^{\rm ND}(x,x')$.  
All Green's functions, representing $G(x,x')$, are expanded in multipoles, 
in terms of the associated Legendre functions $P_\ell^m(\cos\theta)$ 
in spherical coordinates $(r,\theta,\phi)$, 
\beqn
G(x,x')
&=&\sum_{\ell=0}^{\infty}
g_\ell(r,r')
\sum_{m=0}^{\ell}\epsilon_m \frac{(\ell-m)!}{(\ell+m)!}
\nonumber \\
&&\!\!\!\!\!\!\!\!\!\!\!\!\!\!\!
\times
P_\ell^m(\cos\theta)
P_\ell^m(\cos\theta')
\cos m(\phi-\phi'),\ 
\label{eq:Greenfn}
\eeqn
where the coefficient $\epsilon_m$ is defined by 
\beq
\epsilon_m=
\left\{
\begin{array}{cc}
1, & \mbox{ for } m = 0, \\
2, & \mbox{ for } m \geq 1, 
\end{array}
\right.
\eeq
and hence the difference appears in the radial part of the 
Green's function $g_\ell(r,r')$, \cite{Jackson} .

\subsection{Green's function without boundary $G^{\rm NB}(x,x')$}

For the Green's function of the Laplacian without boundary 
$G^{\rm NB}(x,x')$, the radial part $g^{\rm NB}_\ell(r,r')$,
is defined by 
\beq
g^{\rm NB}_\ell(r,r'):=\frac{r_<^\ell}{r_>^{\ell+1}}, 
\eeq
where $r_> := \sup\{r,r'\}$, and $r_< := \inf\{r,r'\}$.  

When the Green's function $G^{\rm NB}$ is applied to BHCS with two concentric 
boundary spheres $S_a$ and $S_b$ located at $r=r_a$ and $r_b$, respectively, 
the following values are used to evaluate surface integrals of 
Eq.~(\ref{eq:GreenIde}), 
\beqn
g^{\rm NB}_\ell(r,r_a) &=& \frac{r_a^\ell}{r^{\ell+1}},\\ 
g^{\rm NB}_\ell(r,r_b) &=& \frac{r^\ell}{r_b^{\ell+1}},   
\eeqn
and, since
$\na'^a G^{\rm NB}(x,x')dS'_a=\mp\pa_{r'}G^{\rm NB}\, r'^2dr'd\Omega'$, 
\beqn
\pa_{r'}g^{\rm NB}_\ell(r,r_a) &=& \ell \,\frac{r_a^{\ell-1}}{r^{\ell+1}},
\label{eq:Apdgnb}
\\
\pa_{r'}g^{\rm NB}_\ell(r,r_b) &=& -(\ell+1)\frac{r^\ell}{r_b^{\ell+2}}.  
\eeqn
Note that the form of (\ref{eq:Apdgnb}) indicates that the Green's function $G^{\rm NB}$
does not pick up $\ell=0$ mode of the Dirichlet data at the sphere $S_a$ ($r=r_a$).

\subsection{Green's function for the region between two concentric 
spheres with Dirichlet conditions, $G^{\rm DD}(x,x')$}
\label{sec:APGDD}

The Green's function $G^{\rm DD}(x,x')$ is a solution of Eq. (\ref{eq:Greeneq})
in a region between two concentric spheres $S_a$ and $S_b$ with radius 
$r=r_a$ and $r=r_b$ ($r_a < r_b$) where Dirichlet 
conditions are imposed.  Its radial part $g^{\rm DD}_\ell(r,r')$
associated with the $\ell$th mode is written
\beqn
&&g^{\rm DD}_\ell(r,r') \,=\,
\left[1-\left(\frac{r_a}{r_b}\right)^{2\ell+1}\right]^{-1}
\frac{r_a^\ell}{r_b^{\ell+1}}
\nonumber\\
&&\!\!\!\!
\times
\left[
\left(\frac{r_{<}}{r_a}\right)^{\ell} 
-\left(\frac{r_a}{r_{<}}\right)^{\ell+1}
\right]
\left[
\left(\frac{r_b}{r_{>}}\right)^{\ell+1}
-\left(\frac{r_{>}}{r_b}\right)^{\ell}
\right].
\label{eq:GreenDDrad}
\eeqn

By construction $g^{\rm DD}_{\ell}(r,r')$ vanishes on the two spheres $S_a$ and $S_b$, 
\beqn
g^{\rm DD}_\ell(r,r_a)&=&0,\\
g^{\rm DD}_\ell(r,r_b)&=&0.  
\eeqn
The derivatives that are used to compute the surface integral in Eq.~(\ref{eq:GreenIde}), 
are
\beqn
\pa_{r'}g^{\rm DD}_\ell(r,r_a)
&=&
\left[1-\left(\frac{r_a}{r_b}\right)^{2\ell+1}\right]^{-1}
\nonumber\\
&&\!\!\!\!\!\!\!\!\!\!\!\!\!\!\!\!\!\!\!\!\!\!\!\!\!\!
\times
(2\ell+1)\,\frac{r_a^{\ell-1}}{r_b^{\ell+1}}
\left[\left(\frac{r_b}{r}\right)^{\ell+1}
-\left(\frac{r}{r_b}\right)^{\ell}\right],
\\
\pa_{r'}g^{\rm DD}_\ell(r,r_b)
&=&
-\left[1-\left(\frac{r_a}{r_b}\right)^{2\ell+1}\right]^{-1}
\nonumber\\
&&\!\!\!\!\!\!\!\!\!\!\!\!\!\!\!\!\!\!\!\!\!\!\!\!\!\!
\times
(2\ell+1)\frac{r_a^\ell}{r_b^{\ell+2}}
\left[\left(\frac{r}{r_a}\right)^\ell
-\left(\frac{r_a}{r}\right)^{\ell+1}\right], 
\eeqn
at $S_a$ and $S_b$, respectively.

\subsection{Green's function for the region between two concentric 
spheres with Neumann and Dirichlet conditions, $G^{\rm ND}(x,x')$}

Similarly, $G^{\rm ND}(x,x')$, is the Green's function between 
$S_a$ and $S_b$, where Neumann data are imposed on $S_a$ and 
Dirichlet data are imposed on $S_b$.  
Its radial part $g^{\rm ND}_\ell(r,r')$
associated with the $\ell$th mode can be written
\beqn
g^{\rm ND}_\ell(r,r')&=&
\left[
1+\frac{\ell}{\ell+1}
\left(\frac{r_a}{r_b}\right)^{2\ell+1}\right]^{-1}
\frac{r_a^\ell}{r_b^{\ell+1}}
\nonumber\\
&&
\times
\left[
\left(\frac{r_<}{r_a}\right)^\ell 
+ \frac{\ell}{\ell+1} \left(\frac{r_a}{r_<}\right)^{\ell+1}
\right]
\nonumber\\
&&
\times
\left[
\left(\frac{r_b}{r_>}\right)^{\ell+1}
- \left(\frac{r_>}{r_b}\right)^{\ell}
\right] .
\label{eq:GreenNDrad}
\eeqn

The values of $g^{\rm ND}_\ell$ 
at the surfaces $S_a$ and $S_b$ become 
\beqn
g^{\rm ND}_\ell(r,r_a)&=&
\left[
1+\frac{\ell}{\ell+1}
\left(\frac{r_a}{r_b}\right)^{2\ell+1}\right]^{-1}
\nonumber\\
&&\!\!\!\!\!\!\!\!\!\!\!\!\!\!\!\!\!\!\!\!
\times
\frac{2\ell+1}{\ell+1}
\frac{r_a^\ell}{r_b^{\ell+1}}
\left[
\left(\frac{r_b}{r}\right)^{\ell+1}
- \left(\frac{r}{r_b}\right)^{\ell}
\right],
\\
g^{\rm ND}_\ell(r,r_b)&=&0, 
\eeqn
and its radial derivatives,  
\beqn
\pa_{r'}g^{\rm ND}_\ell(r,r_a)
&=& 0
\\
\pa_{r'}g^{\rm ND}_\ell(r,r_b)
&=&
-\left[
1+\frac{\ell}{\ell+1}
\left(\frac{r_a}{r_b}\right)^{2\ell+1}\right]^{-1}
\nonumber\\
&&\!\!\!\!\!\!\!\!\!\!\!\!\!\!\!\!\!\!\!\!\!\!\!\!\!\!\!\!\!\!\!\!\!\!\!\!\!\!\!\!\!\!\!\!
\times
(2\ell+1)
\frac{r_a^\ell}{r_b^{\ell+2}}
\left[
\left(\frac{r}{r_a}\right)^\ell 
+ \frac{\ell}{\ell+1} \left(\frac{r_a}{r}\right)^{\ell+1}
\right].
\eeqn
 
The values of $\pa_{r'}g^{\rm ND}_\ell(r,r_a)$ and 
$g^{\rm ND}_\ell(r,r_b)$ vanish by construction.

\section{3+1 decomposition for the Einstein equations}
\label{sec:eqs}

In the usual (3+1) decomposition of the Einstein equations the spacetime metric is written as
\begin{eqnarray}
ds^{2} & = & g_{\GA\GB}dx^{\GA}dx^{\GB} \nonumber \\
       & = & -\GA^{2}dt^{2}+\GG_{ij} (dx^{i}+\GB^{i}dt) (dx^{j}+\GB^{j}dt)                   
\label{metric}
\end{eqnarray} 
where $g_{\GA\GB}$, $\GG_{ij}$, are the 4D and 3D metrics, while $\GA$ and $\GB^{i}$, are the
lapse scalar and the shift vector respectively. The Riemannian 3-metric $\GG_{ij}$ on a 
hypersurface $\Sigma$ is identified by the 4-tensor 
\be
\GG_{\GA\GB}=g_{\GA\GB}+n_{\GA}n_{\GB}
\label{gamma}
\ee 
where $n_{\GA}=-\alpha \nabla_{\GA}t $ is the unit future pointing normal to the hypersurface $\Sigma$.
The indices of $\GG_{\GA\GB}$ can be raised either by $\GG^{\GA\GB}$ or by the full metric 
$g^{\GA\GB}$ and that $\GG^{\GA}_{\:\:\GB}$ projects vectors onto the subspace orthogonal to $n^{\GA}$.
Note that $g_{\GA\GB}$ and $\GG_{\GA\GB}$ differ only on the time-time component while $g^{\GA\GB}$ and 
$\GG^{\GA\GB}$ have identical only the space-space components ($g^{\GA\GB}g_{\GB\GG}=\GD^{\GA}_{\:\: \GG}, \:\:\:
\GG^{ij}\GG_{jk}=\GD^{i}_{\:\: j}, \:\:\: \GG^{\GA\GB}\GG_{\GB\GG} \neq \GD^{\GA}_{\:\: \GG}$). The 
covariant components of the shift are $\GB_{j}=\GG_{ij} \GB^{i}$ and the components on the normal vector are
\be
n_{\GA}=(\GA,0,0,0) \:\:\:\:\: \textup{and} \:\:\:\:\: n^{\GA}=(\frac{1}{\GA},\frac{-\GB^{i}}{\GA})
\label{unitnormalcomponents}
\ee 
The extrinsic curvature is 
\be
K_{\GA\GB}=-D_{\GA}n_{\GB}=-\GG_{\GA}^{\: \GA'}\GG_{\GB}^{\: \GB'} \: \nabla_{\GA'} n_{\GB'}
          =- \frac{1}{2} \pounds_{\mathbf{n}} \GG_{\GA\GB}
\label{extrinsic}
\ee 
where $\nabla$, $D$ are the covariant derivatives associated with $g_{\GA\GB}$ and $\GG_{\GA\GB}$
respectively. The Einstein equations can now be split into the constraint equations
\begin{eqnarray}
\mathcal{R}-K_{ij} K^{ij}+K^{2}=16\pi\GR  \label{ham2}   \\
D_{j}(K^{ij}-\GG^{ij}K) = 8\pi j^{i}      \label{mom2} 
\end{eqnarray} 
and the evolution equations
\begin{eqnarray}
\frac{\OO \GG_{ij}}{\OO t}=-2\GA K_{ij}+2D_{i}\GB_{j}+2D_{j}\GB_{i} \label{evol1}   \:\:\:\:\:\:\:\:\:\:\:\:\: \\
\frac{\OO K_{ij}}{\OO t}=\GA \mathcal{R}_{ij} - D_{i}D_{j} \GA +\GA (KK_{ij}-2K_{im}K_{j}^{\:\: m})  \nonumber \\
   + K_{mi}D_{j} \GB^{m} + K_{mj}D_{i} \GB^{m} + \GB^{m} D_{m}K_{ij}  \label{evol2}  \\
   -8 \pi \GA \left( T_{ij}+\frac{1}{2}\GG_{ij}(\GR-T_{m}^{\:\: m}) \right)    \nonumber
\end{eqnarray} 
where $\GR=T_{\GA\GB} n^{\GA} n^{\GB}$ and $j^{\GA}=-T_{\GB\GG} n^{\GG} \GG^{\GA\GB}$ are the energy and 
momentum density respectively as seen by an observer with four velocity $n^{\GA}$ while $\mathcal{R}_{ij}$, 
$\mathcal{R}$, are the three dimensional Ricci tensor and Ricci scalar on the hypersurface $\Sigma$. From
the two evolution equations we can find the time derivative of the trace of the extrinsic curvature 
\be
\OO_{t} K=\GA \mathcal{R} - \triangle{\GA} +\GA K^{2}+\GB_{i}D^{i}K-8\pi \GG^{ij}P_{ij}
\label{tracetimederivative}
\ee
where $P_{ij}=T_{ij}+\frac{1}{2}\GG_{ij}(\GR-T_{m}^{\:\: m})$ are the source terms and 
$\triangle=D^{i} D_{i}$. 

With a conformal transformation of the form 
\be
\GG_{ij}=\GC^{\GL} \: \TDD{\GG}{i}{j}
\label{conformalgamma}
\ee
the Ricci tensor becomes
\begin{eqnarray}
R_{ij} & = & \frac{\GL( \GL+2)}{4 \GC^{2}}\TD{D}{i}\GC \TD{D}{j}\GC - 
             \frac{\GL( \GL-2)}{4 \GC^{2}}\TDD{\GG}{i}{j}\TU{D}{m}\GC \TD{D}{m}\GC    \nonumber \\
       &   & -\frac{\GL}{2 \GC} (\TD{D}{i}\TD{D}{j}\GC+\TDD{\GG}{i}{j} \TU{D}{m}\TD{D}{m} \GC) 
             +\tilde{\mathcal{R}}_{ij}  
\label{riccitensortrans}
\end{eqnarray}
and the Ricci scalar
\be
R=\GC^{- \GL}\tilde{\mathcal{R}}-\frac{\GL( \GL-4)}{2 \GC^{\GL+2}}\TU{D}{i}\GC \TD{D}{i}\GC -
       \frac{2 \GL}{\GC^{\GL+1}} \TU{D}{m}\TD{D}{m} \GC  \:.
\label{ricciscalartrans}
\ee
Now the Hamiltonian (\ref{ham2}) and the momentum constraint (\ref{mom2}) can be written as 
\begin{eqnarray}
\tilde{\triangle} \GC -\frac{\GC}{2\GL}\tilde{\mathcal{R}}+\frac{\GL-4}{4\GC}\TD{D}{i}\GC \TU{D}{i} \GC
    +\frac{\GC^{\GL+1}}{2\GL}K_{ij}K^{ij}                              \nonumber \\
    - \frac{\GC^{\GL+1}}{2\GL}K^{2} = -\frac{8\pi\GR\GC^{\GL+1}}{\GL}  \label{ham3}  \\
\TD{D}{j}K^{ij}+\frac{5\GL}{2\GC}K^{ij}\TD{D}{j}\GC
               -\frac{\GL}{2\GC^{\GL+1}}K\TUU{\GG}{i}{j}\TD{D}{j}\GC        \nonumber \\
               -\frac{1}{\GC^{\GL}}\TUU{\GG}{i}{j}\TD{D}{j}K = 8 \pi j^{i}  \:. \label{mom3}  
\end{eqnarray}
$\TD{D}{k}$ is the covariant derivative with respect to $\TDD{\GG}{i}{j}$ and 
$\tilde{\triangle}=\TU{D}{i} \TD{D}{i}$. Note also that 
$D_{i}\GO_{j} = \TD{D}{i} \GO_{j} - C^{m}_{\:\: ij} \GO_{m} $ and 
$ C^{m}_{\:\: ij}=\displaystyle \frac{\GL}{2 \GC} (\GD^{m}_{\:\: i} \TD{D}{j}\GC 
  + \GD^{m}_{\:\: j} \TD{D}{i}\GC-\TDD{\GG}{i}{j}\TUU{\GG}{m}{k}\TD{D}{k}\GC) $.

\subsection{Field equations for the initial data}

Since we are searching for quasi-equilibrium  states we assume the existence of a Killing vector
\be
\xi^{\GA}=t^{\GA}+\GZ^{\GA}= \left( \frac{\OO}{\OO t} \right)^{\GA}
          +\Omega \left( \frac{\OO}{\OO \GP} \right)^{\GA} = (1,\GZ^{i}) \label{helicalvector}
\ee
where $\GZ^{i}=\Omega(\OO / \OO \GP)^{i}$, and $\Omega$ is a constant representing the orbital angular 
velocity. In the presence of $\xi^{\GA}$ the spatial metric $\GG_{ij}$ and the extrinsic curvature 
$K_{ij}$ satisfy
\be
\pounds_{\xi} \GG_{ij} = 0  \:\:\:\:\:\:\:\:\: \textup{and} \:\:\:\:\:\:\:\:\: 
\pounds_{\xi} K_{ij}   = 0  \: . \label{helicalsymmetries}
\ee
From the first equation of (\ref{helicalsymmetries}) we get 
\be
\OO_{t}\GG_{ij}+D_{i}\GZ_{j}+D_{j}\GZ_{i}=0 
\ee
thus the time derivative of the spatial metric is associated with the spatial derivative of the rotational 
vector $\GZ^{i}$. With the help of the evolution equation of $\GG_{ij}$, (\ref{evol1}) we can obtain an 
expression for the extrinsic curvature by eliminating the time derivative of the three metric and therefore 
cast the initial value equations in a form that has no time derivatives. We find
\be
K_{ij}=\frac{1}{2\GA} (D_{i}\GO_{j}+D_{j}\GO_{i})  \:\:\:\:\:\:\: \textup{and} \:\:\:\:\:\:\: 
K=\frac{1}{\GA}D_{i}\GO^{i}  
\label{KijK}
\ee 
where $\GO_{i}=\GB_{i}+\GZ_{i}$ is the comoving shift. Since
\be
K_{ij}K^{ij}=\frac{1}{4\GA^{2}}(\LL \GO)_{ij}(\LL \GO)^{ij}+\frac{1}{3}K^{2}
\ee
where
\be
(\LL \GO)_{ij}=D_{i}\GO_{j}+D_{j}\GO_{i}-\frac{2}{3}\GG_{ij}D_{m}\GO^{m} 
\ee
and $(\LL \GO)_{ij}(\LL \GO)^{ij}=(\tilde{\LL} \tilde{\GO})_{ij}(\tilde{\LL} \tilde{\GO})^{ij}$ the hamiltonian
(\ref{ham3}) and momentum constraints (\ref{mom3}) are written
\begin{eqnarray}
\tilde{\triangle} \GC = \frac{\GC}{2\GL}\tilde{\mathcal{R}}-\frac{\GL-4}{4\GC}\TD{D}{i}\GC \TU{D}{i} \GC 
    -\frac{\GC^{\GL+1}}{8\GL\GA^{2}}\TOD{i}{j}\TOU{i}{j}  \:\:\:\:\:  \nonumber \\
    + \frac{\GC^{\GL+1}}{3\GL}K^{2}-\frac{8\pi\GR\GC^{\GL+1}}{\GL} \:\:\:\:\:\:\:\:\:\:\:\:\:\:\:\:
    \:\:\:\:\:\:\:\:\:\:\:\:\:\:\:\:\:\:\:\:\:\:\:\:\:\:\:\:\:\:   \label{ham4}   \\
\tilde{\triangle} \TU{\GO}{i} = -\frac{1}{3}\TU{D}{i}\TD{D}{j}\TU{\GO}{j}-\tilde{\mathcal{R}}^{i}_{\:\: j}\TU{\GO}{j}
                                +\TD{D}{j}\ln \left(\frac{\GA}{\GC^{3\GL /2}} \right)\TOU{i}{j}   \nonumber  \\
                                +\frac{4\GA}{3}\TU{D}{i}K+16\pi\GA\GC^{\GL} j^{i} \:\:\:\:\:\:\:\:\:\:\:\:\:\:\:\:
    \:\:\:\:\:\:\:\:\:\:\:\:\:\:\:\:\:\:\:\:\:\:\:\:\:\:\:\:\:\:  \label{mom4}
\end{eqnarray} 
Also by using the fact that 
\be
\triangle \GA=\GC^{-\GL} \left( \tilde{\triangle}\GA+\frac{\GL}{2\GC}\TD{D}{i}\GC \TU{D}{i}\GA \right)
\ee
we can rewrite equation (\ref{tracetimederivative}) in the conformal geometry as
\begin{eqnarray}
\tilde{\triangle}\GA & = & -\frac{\GL}{2\GC}\TD{D}{i}\GC \TU{D}{i}\GA
                           +\GC^{\GL} \left( \frac{1}{4\GA}\TOD{i}{j}\TOU{i}{j}\right.  \nonumber    \\ 
                     &   & \left. +\frac{\GA K^{2}}{3}+\TU{\GO}{i}\TD{D}{i}K +4\pi\GA (\GR+T) \right)  
\end{eqnarray}
For the binary black hole case, the sources $\GR$, $T_{ij}$ and $j^{i}$ vanish, and take
$\GL=4$. Under these assumptions, our system of equations is
\begin{eqnarray}
\tilde{\triangle} \GC & = & \frac{\GC}{8}\tilde{\mathcal{R}}
                            -\frac{\GC^{5}}{32\GA^{2}}\TOD{i}{j}\TOU{i}{j}+ \frac{\GC^{5}}{12}K^{2}   
                            \label{PsiEq}  \\
\tilde{\triangle}\GA  & = & \GC^{4} \left( \frac{1}{4\GA}\TOD{i}{j}\TOU{i}{j}+\frac{\GA K^{2}}{3}
                           +\TU{\GO}{i}\TD{D}{i}K  \right)   \:  \nonumber \\
                      &   & -\frac{2}{\GC}\TD{D}{i}\GC \TU{D}{i}\GA  \label{AlphaEq}  \\
\tilde{\triangle} \TD{\GO}{i} & = & -\frac{1}{3}\TD{D}{i}\TD{D}{j}\TU{\GO}{j}
                                    -\tilde{\mathcal{R}}_{ij}\TU{\GO}{j}
                                    +\TU{D}{j}\ln \left(\frac{\GA}{\GC^{6}} \right)\TOD{i}{j}  \nonumber \\
                              &   & +\frac{4\GA}{3}\TD{D}{i}K   \:\:\:\:\:\:\:\:\:\:\:\:\:  \label{OmegaEq1} \\
                              & = & -\tilde{\mathcal{R}}_{ij}\TU{\GO}{j}
                                    +\TU{D}{j}\ln \left(\frac{\GA}{\GC^{6}} \right)\TOD{i}{j}  \nonumber  \\
                              &   & +\TD{D}{i}\left( \frac{2}{\GC}\TU{\GO}{j}\TD{D}{j}\GC \right)
                                    -\frac{K}{3}\TD{D}{i}\GA+\GA\TD{D}{i}K     \label{OmegaEq2}  \\
                              & = & -\tilde{\mathcal{R}}_{ij}\TU{\GO}{j}
                                    +\TU{D}{j}\ln \left(\frac{\GA}{\GC^{6}} \right)\TOD{i}{j}  \nonumber  \\
                              &   & +\TD{D}{i}\left( \frac{8}{\GC}\TU{\GO}{j}\TD{D}{j}\GC \right) 
                                    -\frac{4K}{3}\TD{D}{i}\GA+\TD{D}{i}\TD{D}{j}\TU{\GO}{j}  \label{OmegaEq3}
\end{eqnarray}
In the momentum constraint, the last two expressions, equations (\ref{OmegaEq2}) and (\ref{OmegaEq3}), come from 
the fact that $D_{i}K$ involves the second derivative of the comoving shift $\TD{\GO}{i}$ as follows
\be
\TD{D}{i}\TD{D}{j}\TU{\GO}{j}=\GA\TD{D}{i}K+K\TD{D}{i}\GA
   -\TD{D}{i}\left( \frac{6}{\GC}\TU{\GO}{j}\TD{D}{j}\GC  \right)
\ee

\section{Toy model for improvement of convergence by overlap region}
\label{sec:Aptoy}

\begin{figure}
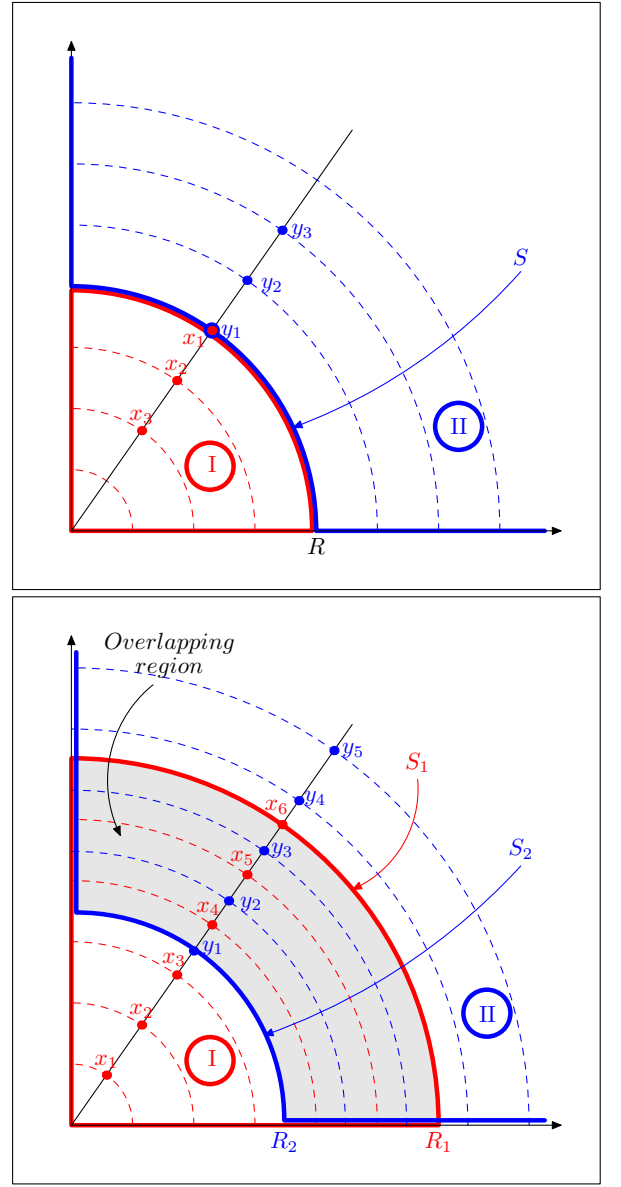

  \includegraphics[width=3.1in,clip]{figure1_2.2.ps}
  \includegraphics[width=3.1in,clip]{figure1_2.1.ps}
\caption{Top, solving the poisson equation on two non overlapping grids.
Bottom, solving the poisson equation by using two overlapping grids. }
  \label{fig:overlap}
\end{figure}

To analyze the improvement of the rate of convergence achieved by  
the overlap region, we 
consider a simple model to calculate the potential of a point mass $M$ 
using two overlapping concentric spherical grids as shown in
Fig.~\ref{fig:overlap}, bottom panel.  
The first grid extends from the coordinate center $(r=0)$ to the surface $S_1$, 
and the second grid from $S_2$ to infinity (or practically to a large 
distance).  In Fig.~\ref{fig:overlap}, top panel, there is no overlapping
region, $S_1=S_2=:S$. The potential in region II $\Phi_{\rm II}$, can be 
calculated from the surface integral of the interface $S_2$, 
as 
\beq
\Phi_{\rm II}=\frac{M+m}{r}, 
\eeq
where $m$ is the error inherited from the initial guess for $\Phi$ 
given at the boundary $S_2$.
The potential of the region I $\Phi_{\rm I}$, is a sum of a volume and 
a surface integral at $S_1$, 
\beq
\Phi_{\rm I}=\frac{M}{r}+e.
\eeq  
Again $e$ is the error inherited from the initial guess for $\Phi$ at $S_1$.  
From continuity of potentials at $S_1$, $\Phi_{\rm I}(R_1)=\Phi_{\rm II}(R_1)$, 
therefore 
\beqn
e = \frac{m}{R_1},  
\\
\Phi_{\rm I}=\frac{M}{r}+\frac{m}{R_1}.  
\eeqn
The next step of iteration is to use this value to 
fix $\Phi_{\rm II}$.  In region II, the potential 
at $R_2$ must satisfy $\Phi_{\rm II}(R_2) = \Phi_{\rm I}(R_2)$, 
whence
\beq
\Phi_{\rm II} = \frac{M + m\frac{R_2}{R_1}}{r}.  
\eeq
Iterating this procedure $n$ times, we will have 
\beq
M^{(n)} = M + m\left(\frac{R_2}{R_1}\right)^n, 
\eeq
namely, the error after the n-th iteration is 
$m(R_2/R_1)^n$.  

Analogously for the $\ell$-th multipole component, 
writing the actual value as $A$, and 
the error in region II as $a$, 
\beq
\Phi_{\rm II} = \frac{A+a}{r^{\ell+1}}.  
\eeq
The solution to region I will be 
\beq
\Phi_{\rm I} = \frac{A}{r^{\ell+1}}+ B r^\ell, 
\eeq
where the second term is again an error.  
The boundary condition at $R_1$, 
$\Phi_{\rm I}(R_1) = \Phi_{\rm II}(R_1)$ yields 
\beqn
B &=& \frac{a}{R_1^{2\ell+1}}, 
\eeqn
and from continuity at $R_2$, 
$\Phi_{\rm II}(R_2) = \Phi_{\rm I}(R_2)$, 
\beq
\Phi_{\rm II} = \frac{A + a\left(\frac{R_2}{R_1}\right)^{2\ell+1}}{r}.  
\eeq
Hence the higher multipole converges faster after the n-th iteration, 
\beq
A^{(n)} = A + a \left(\frac{R_2}{R_1}\right)^{n(2\ell+1)}.  
\eeq

The boundaries $S_1$ and $S_2$ are used to communicate 
the information of the physical boundary conditions imposed 
at the asymptotic region, and the inner excision boundary, as well as 
the source from each region to the other.  
At the n-th step of iteration, the values of $\Phi(x')$ and 
$\OO \Phi(x')/ \OO r'$ in the surface integrals in Eq.~(\ref{eq:GreenIde}) 
are calculated from the potential of $(n-1)$ step of the iteration.

It is possible to achieve communication between the two regions without the 
overlap region, by mixing the values of the potentials $\Phi_{\rm I}$ and
$\Phi_{\rm II}$  of the two regions. This is done when we calculate the 
value of $\OO \Phi(x')/ \OO r'$ at the boundary as shown in the top panel 
of Fig.~\ref{fig:overlap}, choosing say the values at grids 
$x_{3}, \: x_{2}, \: y_{1}, \: y_{2}, \: y_{3}$. 
In this way convergence to a correct solution is again obtained, but the 
number of iterations increases approximately ten times, even for simple 
toy problems presented in Sec.\ref{sec:codetest}.  

In the actual binary calculation shown in Fig.~\ref{fig:PE}, 
when the values of $\Phi$ and $\OO \Phi(x')/ \OO r'$ on $S_{o_{1}}$, 
are intepolated from CCS, 
some of these points do not belong to the computational domain 
of CCS. The smaller the overlapping region the more of these points exist.  
(For example at the point $A$ in Fig.~\ref{fig:PE},
when we interpolate the nearby CCS points we find that  
the point in the lower left corner does not belong to CCS.)
In such case we interpolate the nearby BHCS points to find the value of the
potential.

\section{Convergence of the iteration}
\label{sec:APconv}

In section \ref{sec:Poisol} we have said that a requisit of the Green's function for the KEH iteration to achieve convergence
is that all multipole components of $\nabla'^a G(x,x')$ should not vanish at the boundary for solving Dirichlet problem, 
and similary $G(x,x')$ should not vanish for Neumann boundary conditions. In this Appendix we use the fixed point theorem
to illustrate the convergence (or not) of the KEH method for simple spherically symmetric case with the flat Green's function.

Let's start with the following problem:
\be
\nabla^{2}\varphi =0\: , \:\: r\geq r_a  \:\:\:\: \textrm{with} \:\:\:\: 
                     \frac{\OO \varphi}{\OO r}+\frac{\varphi}{2r}=0 \:\:\:\: \textrm{at} \:\:\: r=r_{a},
\ee
and $\lim_{r\rightarrow \infty}\varphi (r)=1$, whose solution is $\varphi=1+r_{a}/r$.
If we consider the map defined by
\be
\Phi (\varphi)= \frac{1}{4\pi} \int_{S_a} \left[G(x,x') \frac{\OO \varphi}{\OO n'} - \varphi(x')\frac{\OO G}{\OO n'} \right]dS'
\label{Functional}
\ee
on a suitable Sobolev space, where $S_a$ is the surface $r=r_{a}$, we can show the following:

\noindent \underline{claim:} If $G(x,x')=\frac{1}{|x-x'|}$ and the functions
$\varphi(r)$ satisfy 
$\frac{\OO \varphi}{\OO r}+\frac{\varphi}{2r}=0$ on $S_a$ then $\Phi$ has a fixed point. \\
Indeed if $\varphi_{1}$ and $\varphi_{2}$ satisfy the above boundary condition so is $\varphi_{1}-\varphi_{2}$ therefore 
\begin{eqnarray}
\textrm{d}(\Phi(\varphi_{1}), \Phi(\varphi_{2})) & = & \sup_{r} \{ |\Phi(\varphi_{1})-\Phi(\varphi_{2})| \}  \nonumber \\
                                        & \leq & \frac{1}{2} \textrm{d}(\varphi_{1}, \varphi_{2})
\end{eqnarray}
where for the term with $\frac{\OO (\varphi_{1}-\varphi_{2})}{\OO n'}$ we used the boundary condition and for the term with 
$\varphi_{1}-\varphi_{2}$ we used the fact that it depends only on $r$ thus can be pulled out of the integral which when calculated
gives zero. Therefore $\Phi$ has a fixed point which can be found if we take an initial value $\varphi_{0}$ and then compute the 
sequence $\{ \varphi_{n} \}$ with $\varphi_{n+1}=\Phi (\varphi_{n})$ (KEH method). By doing so and adding the contibution from infinity 
which is $1.0$ we get $\varphi_{n+1}(r)=1+\frac{r_{a} \varphi_{n}(r_{a})}{2r}$ which tends to $1+\frac{r_{a}}{r}$ as $n\rightarrow \infty$.

Now if we change the boundary condition to $\varphi=0$ at $r=r_{a}$ the solution turns out to be $\varphi=1-r_{a}/r$. In this case the above 
argument breaks down and as we have seen the KEH iteration fails. The above argument gives us
\be
\textrm{d}(\Phi(\varphi_{1}), \Phi(\varphi_{2})) \leq r_{a} \left| \frac{\OO (\varphi_{1}-\varphi_{2})}{\OO r} \right|_{r=r_{a}} \:.
\ee 
and nothing guaranties that $\Phi$ will have a fixed point any more.
Actually $\varphi_{n+1}(r)=1 - \frac{r_{a}^{2}}{r} \left(\frac{d\varphi_{n}}{dr}\right)_{r=r_{a}}$ thus starting from any constant value, the 
sequence is stuck at $1.0$ and this explains why our code gives everywhere the value $1.0$ with the Dirichlet boundary condition.

\end{document}

%% file: kigou.tex
\newcommand{\beq}{\begin{equation}} 
\newcommand{\eeq}{\end{equation}} 
\newcommand{\beqn}{\begin{eqnarray}} 
\newcommand{\eeqn}{\end{eqnarray}} 
\newcommand{\pa}{\partial}
\newcommand{\na}{\nabla}

\newcommand{\gmabu}{\gamma^{ab}}
\newcommand{\gmabd}{\gamma_{ab}}
\newcommand{\tgmabu}{\tilde\gamma^{ab}}
\newcommand{\tgmabd}{\tilde\gamma_{ab}}

\newcommand{\albe}{{\alpha\beta}}

\newcommand{\Gabd}{G_{\alpha\beta}}

\newcommand{\tD}{\tilde D}
\newcommand{\zD}{{\raise1.0ex\hbox{${}^{\ \circ}$}}\!\!\!\!\!D}
\newcommand{\alone}{{\raise0.5ex\hbox{${}^{\ 1}$}}\!\!\!\!\alpha}

\newcommand{\dl}{\delta}

\newcommand{\Dl}{\Delta}

\newcommand{\Lie}{\mbox{\pounds}}

\newcommand{\nalam}{\mathrel{\raise0.9ex\hbox{$^\lambda$}\mkern-14mu
\lower0.0ex\hbox{$\nabla$}}}

\newcommand{\zeroD}{{\raise1.0ex\hbox{${}^{\ \circ}$}}\!\!\!\!\!D}
\newcommand{\zLap}{{\raise1.0ex\hbox{${}^{\ \circ}$}}\!\!\!\!\Delta}
\newcommand{\zna}{{\raise1.0ex\hbox{${}^{\ \circ}$}}\!\!\!\!\!\nabla}
\newcommand{\tLap}{{\tilde \Delta}}


%% file: final.bbl
\begin{thebibliography}{99}
\bibitem{BTJ2004}
B. Brügmann, W. Tichy, and N. Jansen
Phys. Rev. Lett. 92, 211101 (2004)  
\bibitem{ABDGo2005} M. Alcubierre et al., Phys. Rev. D. \textbf{72}, 044004 (2005).
\bibitem{Pretorius2005} F. Pretorius,  Phys. Rev. Lett. \textbf{95}, 121101 (2005). 
\bibitem{NASA2006}
John G. Baker, Joan Centrella, Dae-Il Choi, Michael Koppitz, and James van Meter, 
Phys. Rev. Lett. 96, 111102 (2006);
John G. Baker, Joan Centrella, Dae-Il Choi, Michael Koppitz, and James van Meter, 
Phys. Rev. D 73, 104002 (2006);
James R. van Meter, John G. Baker, Michael Koppitz, and Dae-Il Choi, 
Phys. Rev. D 73, 124011 (2006).
\bibitem{BROWN2006}  M. Campanelli, C. O. Lousto, Y. Zlochower, 
Phys. Rev. D 73, 061501(R) (2006);
M. Campanelli, C. O. Lousto, P. Marronetti, and Y. Zlochower
Phys. Rev. Lett. 96, 111101 (2006).
\bibitem{Diener2006} P. Diener, F. Herrmann, D. Pollney, E. Schnetter, E. Seidel, R. Takahashi, J. Thornburg, J. Ventrella, 
Phys. Rev. Lett. 96, 121101 (2006).
\bibitem{HSL2006}  F. Herrmann, D. Shoemaker, P. Laguna, gr-qc/0601026.
\bibitem{Brug2006} B. Brügmann, J.A. Gonz\'alez, M. Hannam, S. Husa, U. Sperhake, and W. Tichy, preprint gr-qc/0610128.
\bibitem{Baumg2000} T. Baumgarte, Phys. Rev. D. \textbf{62}, 024018 (2000).
\bibitem{PTC2000} H. P. Pfeiffer, S. A. Teukolsky, and G. B. Cook, Phys. Rev. D. \textbf{62}, 104018 (2000).
\bibitem{MM2000} P. Marronetti and R. A. Matzner, 
Phys. Rev. Lett. \textbf{85}, 5500 (2000).
\bibitem{GGB2002} E. Gourgoulhon, P. Grandcl\'{e}ment, and S. Bonazzola, Phys. Rev. D. \textbf{65}, 044020 (2002); 
P. Grandcl\'{e}ment, E. Gourgoulhon, and S. Bonazzola, Phys. Rev. D. \textbf{65}, 044021 (2002). 
\bibitem{Cook2002} G. Cook, Phys. Rev. D. \textbf{65}, 084003 (2002).  
\bibitem{PCT2002} H. Pfeiffer, G. B. Cook, and S. Teukolsky, 
Phys. Rev. D. \textbf{66},  024047 (2002)
\bibitem{TBCD2003}  W. Tichy, B. Br\"{u}gmann, M. Campanelli, P. Diener, 
Phys. Rev. D { \bf 67}, 064008 (2003). 
\bibitem{Ansorg} M. Ansorg, B. Br\"{u}gmann, and W. Tichy, 
Phys. Rev. D { \bf 70}, 064011 (2004); 
M. Ansorg, Phys. Rev. D { \bf 72}, 024018 (2005). 
\bibitem{YCST2004}  Hwei-Jang Yo, J. Cook, S. Shapiro, T. Baumgarte, Phys. Rev. D. 
\textbf{70},084033 (2004); Erratum ibid. \textbf{70} 089904 (2004)
\bibitem{Hannam2005} M. Hannam, Phys. Rev. D. \textbf{72}, 044025 (2005).
\bibitem{Cook2000} G. B. Cook, Living Rev. Relativity \textbf{3}, 1 (2000).
\bibitem{KEH89} H. Komatsu, Y. Eriguchi, and I. Hachisu, MNRAS {\bf 237}, 355 (1989). 
\bibitem{USE00}
  K. Ury\=u and Y. Eriguchi, Phys. Rev. D. {\bf 61}, 124023 (2000); 
  K. Ury\=u, M. Shibata and Y. Eriguchi, Phys. Rev. D. {\bf 62}, 
104015 (2000). 
%
\bibitem{SourcesOfGR} J. W. York, Jr., in \emph{Sources of Gravitational Radiation}, 
edited by L. Smarr (Cambridge, Cambridge, 1979), pp. 83-126.  
%
\bibitem{ISEN78} 
J. Isenberg, 
Waveless Approximation Theories of Gravity, preprint (1978),
University of Maryland;
J. Isenberg and J. Nester, in {\sl General Relativity
and Gravitation} Vol.1, edited by A. Held, (Plenum Press, New York 1980).
\bibitem{WMM959} 
   J. R. Wilson and G. J. Mathews, Phys. Rev. Lett. {\bf 75}, 4161 (1995); 
   P. Marronetti, G. J. Mathews and J. R. Wilson, 
   Phys. Rev. D {\bf 60}, 087301 (1999). 
\bibitem{BA97} 
   T. W. Baumgarte, G. B. Cook, M. A. Scheel, S. L. Shapiro and 
  S. A. Teukolsky, Phys. Rev. D {\bf 57}, 6181 (1998);
  {\it ibid} {\bf 57}, 7299 (1998). 
\bibitem{SUF04} 
M. Shibata, K. Ury\=u, and J. L. Friedman, Phys. Rev. D {\bf 70},
044044 (2004); Erratum ibid. {\bf 70}, 129901(E) (2004).
\bibitem{BY80} 
J.~M. Bowen, and J.~W. York, \prd {\bf 21}, 2047 (1980).
%
\bibitem{BL63} 
D.~R. Brill and R.~W. Lindquist, Phys. Rev. 131, 471 (1963).
%
\bibitem{Yoshida06}
S'i. Yoshida, B.~C. Bromley, J.~S. Read, K. Ury\=u, and J.~L. Friedman, 
Classical Quantum Gravity \textbf{23}, S599 (2006).  
\bibitem{pepe} 
J.L. Jaramillo, E. Gourgoulhon, G.A. Mena Marugan, \prd {\bf 70}, 124036 (2004); 
E. Gourgoulhon, J.L. Jaramillo, Phys.Rept., 423, 159, (2006).
\bibitem{Cook2004} 
G. B. Cook, H. P. Pfeiffer, Phys.Rev. D {\bf 70}, 104016, (2004).
\bibitem{SU06} M. Shibata, K. Ury\=u, Phys. Rev. D \textbf{74}, 121503(R) (2006); 
astro-ph/0611522 [Classical Quantum Gravity (to be published)].
\bibitem{NKO1984} 
T. Nakamura, Y. Kojima, and K.-I. Oohara, Phys. Lett. A, 106, 235, 1984; 
K.-I. Oohara, T. Nakamura, and Y. Kojima, Phys. Lett. A, 107, 452, 1985.
\bibitem{apshi} 
M. Shibata, Phys. Rev. D {\bf 55}, 2002 (1997); 
M. Shibata, and K. Ury\=u, Phys. Rev. D {\bf 62}, 087501 (2000).
\bibitem{Thorn2004} 
J. Thornburg, Class. Quant. Grav., 21, 743 (2004).
\bibitem{Jackson} 
J. D. Jackson, \emph{Classical Electrodynamics}, Second Ed., (John Wiley \& Sons, 1975).
\bibitem{Lin} Lap-Ming Lin and Jerome Novak, gr-qc/0702038.



\end{thebibliography}
